\documentclass[aps,pre,showpacs,twocolumn]{revtex4}%
\usepackage{amsfonts}
\usepackage{amsmath}
\usepackage{amssymb}
\usepackage{graphicx}%
\providecommand{\U}[1]{\protect\rule{.1in}{.1in}}

\begin{document}
\title{Shear viscosity: velocity gradient as a constraint on wave function}
\author{M.-L. Zhang and D. A. Drabold}
\affiliation{Department of Physics and Astronomy, Ohio University, Athens, Ohio 45701}

\begin{abstract}
By viewing a velocity gradient in a fluid as an internal disturbance and
treating it as a constraint on the wave function of a system, a linear
evolution equation for the wave function is obtained from the Lagrange
multiplier method. It allows us to define the microscopic response to a
velocity gradient in a pure state. Taking a spatial coarse-graining average
over this microscopic response and averaging it over admissible initial
states, we achieve the observed macroscopic response and transport
coefficient. In this scheme, temporal coarse-graining is not needed. The
dissipation caused by a velocity gradient depends on the square of initial
occupation probability, whereas the dissipation caused by a mechanical
perturbation depends on the initial occupation probability itself. We apply
the method of variation of constants to solve the time-dependent Schrodinger
equation with constraints. The various time scales appearing in the momentum
transport are estimated. The relation\ between the present work and previous
theories is discussed.

\end{abstract}

\maketitle


\section{Introduction}

\label{intr}

To express macroscopic observables in terms of the microscopic parameters of
composite particles, one has to reduce an infinite number of mechanical
degrees of freedom to a few hydrodynamical degrees of freedom\cite{dgr}. Two
typical techniques are projection and coarse-graining. In the projection
method, one can either project the density matrix to a reduced density matrix
for the relevant degrees of freedom (Schrodinger picture)\cite{zw60,kub} or
project a sum of microscopic mechanical variables to a macroscopic slow
variable (Heisenberg picture)\cite{zw61,sew,mo}. While in the coarse-graining
method, excepting the obligatory ensemble average, one still has to carry out
at least one of three average procedures: (1) coarse-graining in
time\cite{ka1,ori}; (2) coarse-graining in space\cite{rus,bins,bi,jac}; or (3)
coarse-graining in the eigenvalue spectrum of energy and other collective
variables\cite{ka1,sew}.

In the coarse-graining techniques, the spatial coarse grained average plays an
essential role in formulating all macroscopic problems. Macroscopically, a
non-equilibrium process in a fluid is described by some local hydrodynamic
(thermodynamic, macroscopic) variables: temperature, stress, concentration and
macroscopic velocity field etc\cite{dgr}. The fundamental assumption behind
this macroscopic description is that there exists a local equilibrium at each
coarse-grained spatial `point' and at every coarse-grained temporal
`moment'\cite{dgr}. Similarly in the kinetic theories, at least three of the
arguments of the single-particle distribution function, time, position and
translational momentum are defined in a coarse-grained sense (far more
coarse-grained than the requirement of uncertainty principle)\cite{kub,pk}. As
an alternative to the distribution function in phase space, Onsager introduced
a distribution function for the coarse grained moments of the conserved
quantities (displacements)\cite{on1,on2}. To obtain hydrodynamic equations and
transport coefficients, M. S. Green noticed that the resolution of a
macroscopic measurement is finite, so that one must eliminate the high wave
number components of a collective variable which are are unmeasurable. The
observed value of a hydrodynamic variable is given by averaging the
truncated\cite{bins,bi} collective variable over the distribution function of
Markoff process\cite{mel1,mel2}.

It is well-known that temporal coarse-graining (TCG) is not necessary for
deriving the macroscopic Maxwell equations (MAME) from the microscopic
ones\cite{rus,bins,bi}. A safe lower limit of length $L_{0}$ for a macroscopic
description of reflection and refraction of visible light can be taken as
$10^{2}$\AA \cite{jac}. The time scale $t_{0}$ associated with $L_{0}$ is in
the range of atomic or molecular motions, a TCG at a time scale $t_{0}$ is
meaningless\cite{jac}. Moreover, if one averages the motion of the particles
over a time scale longer than $t_{0}$, the scattering phenomena will be
smeared\cite{lv8}. In the MAME, the sources of fields are the total charge
density and the total current density\cite{jac,lv8}. At the same time, the
induced charge density and the induced current density are also the responses
of the system to the external fields. Since no TCG is taken in deriving MAME,
one should not require TCG in computing current density
(conductivity)\cite{short,pss,4t,maw}. The energy dissipation of system is
taken care of by the thermal contact with a thermal bath\cite{cb,maw,jp}. It
seems that TCG is not needed for the irreversible processes caused by the
mechanical distrubances\cite{kyn,k57,rus,bins,bi,jac,lv8,short,pss,4t,maw}.

An open question is whether TCG is necessary to describe an irreversible
process induced by an internal (thermal) disturbance\cite{kyn}. Several
schemes have been designed to replace an internal disturbance with a
fictitious mechanical disturbance\cite{oll,kad,lut,maz,fel}. But these methods
assume \textit{a priori} that the Navier-Stokes equation is valid\cite{zub}.
In other theories, TCG is often taken along with spatial coarse-graining and
ensemble average\cite{kyn,kub,zub,eva,cb,ori}. In particularly, TCG is viewed
as a critical step producing irreversibility in the kinetic approach (master
or Boltzmann equation)\cite{cb,ka1}.

Although a velocity gradient is often viewed as an \textit{internal}
disturbance, it can be realized in a `mechanical' manner. In a fluid, two
obvious ways to produce a velocity gradient are (i) moving a boundary plate
which confines the fluid; and (ii) stirring the fluid with a rod in the
middle. In both situations, a velocity gradient is produced by the interaction
between the fluid and an external object at the solid-fluid interface. A
non-equilibrium `Hamiltonian' has been phenomenologically introduced to
effectively compute viscosity\cite{hoo,ean,eva}. In addition, for a mechanical
disturbance\cite{kyn,k57}, one may define microscopic response in a pure
state. The observed macroscopic response (consequently transport coefficient)
is given by spatial coarse graining the microscopic response and averaging
over possible initial conditions\cite{short,pre,jp,pss}. It is worthwhile to
explore whether we could describe viscosity in a more `mechanical' manner: (i)
not invoking temporal coarse-graining; and (ii) not assuming local equilibrium
as the starting point.\

In this paper, we show that shear viscosity can be strictly calculated from a
time-dependent Schrodinger equation, and time coarse-graining can be avoided.
One may repeat the same procedure for bulk viscosity \textit{mutatis
mutandis}. In the discussion, we will only use the Schrodinger picture.
Because a hydrodynamic quantity is a bilinear form of the many-body wave
function, all the conclusions are valid for any set of identical particles:
bosons or fermions. Only when we take concrete approximations for the
many-body wave functions, do we need to know whether the particles are
fermions or bosons.

In Sec. \ref{cons}, we first express the velocity gradient in a fluid as a
constraint on the wave function of the fluid, cf. Eq.(\ref{bc}). To prescribe
an internal disturbance like velocity gradient, one needs the wave function of
the system. The wave function must be determined self-consistently with the
given velocity gradient. With velocity gradient as a constraint, the evolution
equation for wave function is then derived using the Lagrange multiplier
method\cite{cou}, cf. Eqs.(\ref{eu1},\ref{eu3},\ref{eu4}). From the
appearance, it is a time-dependent Schrodinger equation. Two additional terms
[Eqs.(\ref{eu3},\ref{eu4})] appear in the `Hamiltonian'. Both of them contain
the occupation probabilities of admissible initial states, which reflects the
fact that a velocity gradient is an internal disturbance. On the other hand a
mechanical perturbation is completely specified by the time dependence of
external field\cite{k57}.

In Sec. \ref{shou}, we first discuss the entropy production rate of the system
+ bath, cf. Eqs.(\ref{entr},\ref{tre}). Secondly, we check the mass
conservation law, cf. Eqs.(\ref{hmc0},\ref{hmc}). Thirdly, the microscopic
response Eq.(\ref{flu}) to the velocity gradient (the momentum flux in a pure
state) is obtained from the momentum conservation law in a pure state. The
dissipation caused by a mechanical perturbation depends on the occupation
probabilities of admissible initial states. Because a velocity gradient is an
internal disturbance, the dissipation caused by a velocity gradient depends on
the \textit{squares} of the occupation probabilities of admissible initial
states. This feature is clearly seen in the time rates of change of energy,
mass and momentum.

Representing velocity gradient as a constraint on the wave function
[Eq.(\ref{bc})] depends critically on the assumption that a suitable spatial
coarse-graining is adequate\cite{rus,bins,bi,jac} to describe internal
friction. In Sec. \ref{put} we show that the spatial coarse-graining average
automatically contains a coarse-graining in time [Eq.(\ref{3})] and a
coarse-graining in eigenvalue spectrum of collective variables [Sec.
\ref{egcr}].

In Sec. \ref{dirac}, we apply the method of variation constants to obtain the
Lagrange multipliers which appear in the solution of the time-dependent
Schrodinger equation. The Lagrange multipliers characterize the interaction on
the system exerted by the boundary plates. The phenomenological
non-equilibrium `Hamiltonian'\cite{hoo,ean,eva} can be obtained from $H_{d}$,
the interaction of system with plates linear in velocity gradient. The shear
viscosity is read from the macroscopic momentum flux (\ref{mrf}) which is
deduced from the wave function of system at some moment. In Sec. \ref{tcd}, we
explain that applying Dirac perturbation theory to transport process is valid.
Various time scales in the momentum transport process are discussed. We show
that the length scale of spatial coarse-graining is determined by an intrinsic
time scale.

In Sec. \ref{fph}, we show that the macroscopic stress tensor derived from the
microscopic response method is the same as that derived from averaging the
momentum flux operator over the density matrix. We developed a cumulant
expansion for the density operator. When we replace the operator in the
exponent with its expectation value, we reproduce the non-equilibrium density
matrix obtained from other approaches.

\section{evolution of state driven by velocity gradient}

\label{cons}

A mechanical disturbance on a system can be expressed by the coupling between
the mechanical degrees of freedom of the system and some specified external
parameters which may depend on time and position. The disturbance caused by a
velocity gradient is more complicated. We will see that to describe a fluid
with a velocity gradient, the occupation probabilities of admissible initial
states enter the `Hamiltonian' of the fluid. In Sec.\ref{msm} we briefly
summarize the microscopic response method for a mechanical
perturbation\cite{short,pss,jp,pre}. Later, we will show that one can still
define a momentum flux in a pure state (microscopic response to velocity
gradient), viscosity can be calculated from a modified time-dependent
Schrodinger equation.

\subsection{Macroscopic response to a mechanical disturbance}

\label{msm}

Consider a $N-$particle system $\mathcal{S}$ described by Hamiltonian $H_{0}$,
we use indexes $a,b,\cdots$ to label the eigenvalues $\{E_{a}\}$ and
eigenstates $\{\Phi_{a}\}$ of $H_{0}$:%
\begin{equation}
H_{0}\Phi_{a}=E_{a}\Phi_{a}. \label{eig}%
\end{equation}
If the system $\mathcal{S}$ is in thermal contact with a thermal reservoir
$\mathcal{B}$ at temperature $T$, and is in equilibrium with it\cite{cb,fel},
the probability that the system is in state $\Phi_{b}$ is
\begin{equation}
P^{\Phi_{b}}=e^{-E_{b}/k_{B}T}/\sum_{a}e^{-E_{a}/k_{B}T}. \label{canp}%
\end{equation}

Because the system is macroscopic, comparing the energy of system
$\mathcal{S}$, the energy exchange rate between the system and the bath is
slow\cite{v5}. In a time period much longer than a macroscopic measurement,
the system can be viewed as isolated, and described by a wave function. For a
macroscopic fluid, the detailed dynamics of the surrounding bath is not important.

If a mechanical disturbance described by Hamiltonian $H_{f}$ is exerted on
$\mathcal{S}$, the state $\Psi$ of $\mathcal{S}$ will evolve according to
\begin{equation}
i\hbar\partial\Psi/\partial t=(H_{0}+H_{f})\Psi.\label{tse}%
\end{equation}
The above description requires that\cite{cb,jp} $\mathcal{S}$ is in good
thermal contact with $\mathcal{B}$ such that the energy generated by the
disturbance $H_{f}$ inside $\mathcal{S}$ can be rapidly transferred into
$\mathcal{B}$, and the system is maintained at the temperature $T$ of bath
$\mathcal{B}$. If we assume that the system is initially in state $\Phi_{a}$,
the state $\Psi_{a}(t)$ of system at time $t$ is determined by Eq.(\ref{tse})
with initial condition $\Psi_{a}(t=-\infty)=\Phi_{a}$. For convenience we
adiabatically introduced $H_{f}$. By means of the microscopic conservation law
in a pure state, we can define\cite{short} the corresponding microscopic
response (flux) in state $\Psi_{a}(t)$.

One can always spatially resolve a quantity $A^{\Psi_{a}}(\mathbf{R},t)$ in
state $\Psi_{a}$ into its Fourier components:
\begin{equation}
A^{\Psi_{a}}(\mathbf{R},t)=\int\frac{d^{3}k}{(2\pi)^{3/2}}e^{i\mathbf{k}%
\cdot\mathbf{R}}\mathcal{A}^{\Psi_{a}}(\mathbf{k},t). \label{fr}%
\end{equation}
Because any instrument has finite spatial resolution, one can only detect
those $\mathcal{A}^{\Psi_{a}}(\mathbf{k},t)$ with $|\mathbf{k}|<k_{0}$, where
$k_{0}$ is some instrument limited cut-off wave number\cite{mel2,rus,bi}. To
express a macroscopic measurement, one has to make spatial coarse-grained
average\cite{bi,jac}:%
\begin{equation}
A^{c\Psi_{a}}(\mathbf{r},t)=\int d\mathbf{R}f(\mathbf{r}-\mathbf{R}%
)A^{\Psi_{a}}(\mathbf{R},t), \label{cg}%
\end{equation}
where $f(\mathbf{s})$ is any function satisfying: (1) $\int d\mathbf{s}%
f(\mathbf{s})=1$; and (2) the Fourier components of $f(\mathbf{s})$ tends
rapidly to zero for $|\mathbf{k}|>k_{0}$. The integral in Eq.(\ref{cg}) is
over the sample.

Since the system is in thermal contact with a reservoir, i.e. a member of a
canonical ensemble, one does not know which state the system is initially in.
The measured quantity is an ensemble average of $A^{\Psi_{a}c}(\mathbf{r},t)$:%
\begin{equation}
A(\mathbf{r},t)=\sum_{\Psi_{a}}P^{\Phi_{a}}A^{c\Psi_{a}}(\mathbf{r},t),
\label{ea}%
\end{equation}
where $P^{\Psi_{a}}(-\infty)=P^{\Phi_{a}}$ is the probability that the system
is in state $\Psi_{a}(t=-\infty)=\Phi_{a}$ in the remote past. Since the
microscopic response $A^{\Psi_{a}}(\mathbf{R},t)$ is a bilinear form of
$\Psi_{a}$, it can be computed from Eq.(\ref{tse}) rather than from the
Liouville equation for the density matrix. Since the spatial coarse-graining
(\ref{cg}) and the average over initial conditions (\ref{ea}) do not involve
the temporal evolution of the system, the macroscopic response $A(\mathbf{r}%
,t)$ and consequently transport coefficient are also determined by the
time-dependent Schrodinger equation\cite{cb,short,pss,pre}. The averaging
procedure (\ref{cg},\ref{ea}) is simpler than, and equivalent to Kubo's linear
response theory\cite{pre}.

The procedure (\ref{cg},\ref{ea}) aims at the steady states $\{\Psi_{a}(t)\}$
of $\mathcal{S}$ under a monochromatic driving. The general linear casual
response to a disturbance with arbitrary time dependence will be discussed in
Sec.\ref{caus}.

\subsection{Velocity gradient as a constraint on wave function}

\label{ysu}

Consider a fluid composed of $N$ identical particles with mass $m$. We use
indexes $j$, $k$, $l\cdots$ to label the particles. Denote the interaction
energy between two particles at $\mathbf{r}_{k}$ and $\mathbf{r}_{l}$ as
$U(\mathbf{r}_{k},\mathbf{r}_{l})$. Then the Hamiltonian of system is
\begin{equation}
H_{0}=-\frac{\hbar^{2}}{2m}\sum_{k=1}^{N}\nabla_{\mathbf{r}_{k}}^{2}+\frac
{1}{2}\sum_{k,l(k\neq l)}U(\mathbf{r}_{k},\mathbf{r}_{l}).\label{hs}%
\end{equation}

Put the system in an external field, denote the interaction potential energy
for a particle at $\mathbf{r}_{k}$ in the external field as $V(\mathbf{r}%
_{k},t)$. The interaction Hamiltonian with external field is
\begin{equation}
H_{f}=\sum_{k=1}^{N}V(\mathbf{r}_{k},t). \label{hf}%
\end{equation}

In the remote past $t=-\infty$, suppose that the system is in state $\Phi_{a}%
$. Driving the system with a velocity gradient and $H_{f}$, $\Phi_{a}$ will
evolve into a state $\Psi_{a}(\mathbf{r}_{1}\cdots\mathbf{r}_{N};t)$ at time
$t$. Eq.(\ref{canp}) indicates that we do not have the full knowledge of the
system at the initial moment. The evolution of the system can then be
understood to be probabilistic\cite{tol}. The interaction $U$ among molecules
renders the macroscopic motion involving a velocity gradient
irreversible\cite{tol}.

\subsubsection{Macroscopic velocity field in a fluid}

Using the definition of mass density operator \cite{mel2,kub} $\widehat{\rho
}(\mathbf{r})=\sum_{j=1}^{N}m\delta(\mathbf{r}-\mathbf{r}_{j})$, the mass
density for state $\Psi_{a}(t)$ is a function of time and position:%
\begin{equation}
\rho^{\Psi_{a}}(\mathbf{r},t)=Nm\int d\tau_{1}\Psi_{a}^{\ast}\Psi_{a},
\label{mad}%
\end{equation}
where $d\tau_{1}=d\mathbf{r}_{2}\cdots d\mathbf{r}_{N},$ the arguments of
$\Psi_{a}$ are $(\mathbf{r},\mathbf{r}_{2},\mathbf{r}_{3},\cdots
,\mathbf{r}_{N};t)$. The momentum density operator is defined
as\cite{mel2,kub}%
\begin{equation}
\widehat{p}_{\gamma}(\mathbf{r})=\frac{1}{2}\sum_{j=1}^{N}[-i\hbar
\frac{\partial}{\partial r_{j\gamma}}\delta(\mathbf{r}-\mathbf{r}_{j}%
)+\delta(\mathbf{r}-\mathbf{r}_{j})(-i\hbar\frac{\partial}{\partial
r_{j\gamma}})], \label{dc}%
\end{equation}
where $\gamma=x,y,z$ are indexes for three Cartesian components. The momentum
density in state $\Psi_{a}$ is
\begin{equation}
S_{\gamma}^{\Psi_{a}}(\mathbf{r},t)=\frac{i\hbar}{2}N\int d\tau_{1}(\Psi
_{a}\frac{\partial\Psi_{a}^{\ast}}{\partial r_{\gamma}}-\Psi_{a}^{\ast}%
\frac{\partial\Psi_{a}}{\partial r_{\gamma}}). \label{mod}%
\end{equation}

According to Eqs.(\ref{cg},\ref{ea}), the measured macroscopic mass density is%
\begin{equation}
\rho(\mathbf{r},t)=\sum_{\Psi_{a}}P^{\Phi_{a}}\int d\mathbf{R}f(\mathbf{r}%
-\mathbf{R})Nm\int d\tau_{1}\Psi_{a}\Psi_{a}^{\ast}, \label{md}%
\end{equation}
where the arguments of $\Psi_{a}$ are $(\mathbf{R},\mathbf{r}_{2}%
,\mathbf{r}_{3},\cdots,\mathbf{r}_{N};t)$. The measured macroscopic momentum
density is%
\begin{equation}
S_{\gamma}(\mathbf{r},t)=\sum_{\Psi_{a}}P^{\Phi_{a}}\int d\mathbf{R}%
f(\mathbf{r}-\mathbf{R})S_{\gamma}^{\Psi_{a}}(\mathbf{R},t). \label{ond}%
\end{equation}
The macroscopic velocity field should be defined as\cite{mel2,ori,v6}:%
\begin{equation}
v_{\gamma}(\mathbf{r},t)=S_{\gamma}(\mathbf{r},t)/\rho(\mathbf{r},t),\text{
}\gamma=x,y,z. \label{mfv}%
\end{equation}
It satisfies the requirement that the macroscopic velocity is the macroscopic
momentum of unit mass of fluid\cite{v6}.

\subsubsection{Velocity gradient}

\begin{figure}[ptb]
\begin{center}
\includegraphics[width=8cm]{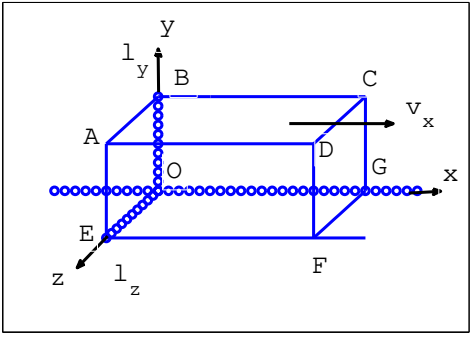}
\end{center}
\caption{Experimental setup for velocity gradient: the top plates ABCD moves
toward $+x$ direction (right) with speed $v_{x0}$. The bottom OEFG, front
(right) bank AEFD and back (left) bank BOGC are static. The width of channel
is $l_{z}$, the depth of channel is $l_{y}$.}%
\label{Fig1}%
\end{figure}

We first describe the simplest experimental setup for measuring shear
viscosity in Fig.\ref{Fig1}. Let the fluid flow in a rectangular channel along
the $x$ direction. We choose a coordinate system such that the two banks BOGC
and AEFD of the channel are $z=0$ and $z=l_{z}$, the bottom surface OEFG of
channel is $y=0$, the top surface ABCD of the fluid is bordered by a plate
$y=l_{y}$ with velocity $v_{x0}$. The points on the bottom plate OEFG are
denoted as $\mathbf{r}^{(1)}(x,0,z)$ with $-\infty<x<\infty$ and $0<z<l_{z}$,
the boundary condition on bottom is $v_{\gamma}(\mathbf{r}^{(1)},t)=0$,
$\gamma=x,y,z$. The points on the top plate ABCD are denoted as $\mathbf{r}%
^{(2)}(x,l_{y},z)$ with $-\infty<x<\infty$ and $0<z<l_{z}$, the boundary
condition on the top plate is $v_{x}(\mathbf{r}^{(2)},t)=v_{x0},v_{y}%
(\mathbf{r}^{(2)},t)=v_{z}(\mathbf{r}^{(2)},t)=0$. The points on the back
(left) bank BOGC of channel are denoted as $\mathbf{r}^{(3)}(x,y,0)$ with
$-\infty<x<\infty$ and $0<y<l_{y}$, the boundary condition on the back bank
BOGC is $v_{\gamma}(\mathbf{r}^{(3)},t)=0$, $\gamma=x,y,z$. The points on the
front (right) bank AEFD of channel are denoted as $\mathbf{r}^{(4)}%
(x,y,l_{z})$ with $-\infty<x<\infty$ and $0<y<l_{y}$, the boundary condition
on the front bank AEFD is $v_{\gamma}(\mathbf{r}^{(4)},t)=0$, $\gamma=x,y,z$.
These no-slip conditions for a viscous fluid are the boundary conditions for
the Navier-Stokes equations\cite{v6}. We imposed a velocity gradient
$v_{x0}/l_{y}$ in the fluid.

\subsubsection{Velocity gradient as constraints on wave function}

To enforce a velocity gradient as a constraint on the wave function, it is
helpful to allow an arbitrary velocity field on each boundary surface. We use
a two dimensional mesh to characterize the positions of the points on a
boundary surface. For a point on the $\sigma^{th}$ plate ($\sigma=1,2,3,4$),
we use two indexes $\mu_{\sigma}$ and $\nu_{\sigma}$ to describe its position,
where $\mu_{\sigma}=1,2,\cdots,M_{1\sigma}$ and $\nu_{\sigma}=1,2,\cdots
,M_{2\sigma}$. Thus the position vector of a point on the $\sigma^{th}$ plate
with indexes $\mu_{\sigma}$ and $\nu_{\sigma}$ is denoted as $\mathbf{r}%
_{\mu_{\sigma}\nu_{\sigma}}^{(\sigma)}$, there are $M_{1\sigma}\times
M_{2\sigma}$ points on the $\sigma^{th}$ boundary surface. A general velocity
field on the $\sigma^{th}$ boundary surface may be specified $v_{\gamma
}(\mathbf{r}_{\mu_{\sigma}\nu_{\sigma}}^{(\sigma)},t)$, $\gamma=x,y,z$. In
terms of Eq.(\ref{mfv}), the boundary condition on the $\sigma^{th}$ boundary
surface is expressed as%
\begin{equation}
S_{\gamma}(\mathbf{r}_{\mu_{\sigma}\nu_{\sigma}}^{(\sigma)},t)-\rho
(\mathbf{r}_{\mu_{\sigma}\nu_{\sigma}}^{(\sigma)},t)v_{\gamma}(\mathbf{r}%
_{\mu_{\sigma}\nu_{\sigma}}^{(\sigma)},t)=0,\label{nsli}%
\end{equation}
where $v_{\gamma}(\mathbf{r}_{\mu_{\sigma}\nu_{\sigma}}^{(\sigma)},t)$ is a
know function of $(\mathbf{r}_{\mu_{\sigma}\nu_{\sigma}}^{(\sigma)},t)$.
Making use of Eqs.(\ref{md},\ref{mod},\ref{ond}), Eq.(\ref{nsli}) becomes
\begin{equation}
\sum_{\Psi_{b}}P^{\Phi_{b}}\int d\mathbf{R}f(\mathbf{r}_{\mu_{\sigma}%
\nu_{\sigma}}^{(\sigma)}-\mathbf{R})N\int d\tau_{1}\label{bc}%
\end{equation}%
\[
\times\lbrack\frac{i\hbar}{2}(\Psi_{b}\frac{\partial\Psi_{b}^{\ast}}{\partial
R_{\gamma}}-\Psi_{b}^{\ast}\frac{\partial\Psi_{b}}{\partial R_{\gamma}%
})-v_{\gamma}(\mathbf{r}_{\mu_{\sigma}\nu_{\sigma}}^{(\sigma)},t)m\Psi_{b}%
\Psi_{b}^{\ast}]=0.
\]
Eq.(\ref{bc}) represents $M_{c}=3\times4\times M_{1\sigma}\times M_{2\sigma}$
constraints on the wave function of system. The evolution equation of
$\Psi_{a}(t)$ has to be modified with constraints (\ref{bc}).

\subsection{Evolution of state in a velocity gradient}

\label{tdu}

To consider the effect of constraints (\ref{bc}) on the state $\Psi_{a}$ of
the many-particle system, we first notice that the Schrodinger equation%
\begin{equation}
i\hbar\Psi_{at}=(H_{0}+H_{f})\Psi_{a} \label{se}%
\end{equation}
is the Euler equation $\delta S/\delta\Psi_{a}^{\ast}=0$ of the
action\cite{cou,if}:%
\begin{equation}
S=\int_{t_{1}}^{t_{2}}dt\int d\tau\mathcal{L}(\Psi_{a},\Psi_{at}%
,\Psi_{a\mathbf{r}_{j}};\Psi_{a}^{\ast},\Psi_{at}^{\ast},\Psi_{a\mathbf{r}%
_{j}}^{\ast}), \label{act}%
\end{equation}
where $\Psi_{at}=\partial\Psi_{a}/\partial t$, $\Psi_{a\mathbf{r}_{j}}%
=\nabla_{\mathbf{r}_{j}}\Psi_{a}$, $d\tau=d\mathbf{r}_{1}d\mathbf{r}_{2}\cdots
d\mathbf{r}_{N}$ is the volume element in configurational space, the arguments
of $\Psi_{a}$ are $(\mathbf{r}_{1},\mathbf{r}_{2},\cdots,\mathbf{r}_{N};t)$.
The well-known Lagrangian density for Eq.(\ref{se}) is\cite{if}:%
\begin{equation}
\mathcal{L}=i\hbar\Psi_{a}^{\ast}\Psi_{at}-\sum_{j=1}^{N}\frac{\hbar^{2}}%
{2m}\Psi_{a\mathbf{r}_{j}}^{\ast}\Psi_{a\mathbf{r}_{j}} \label{lag}%
\end{equation}%
\[
-[\sum_{k=1}^{N}V(\mathbf{r}_{k},t)+\frac{1}{2}\sum_{k,l(k\neq l)}%
U(\mathbf{r}_{k},\mathbf{r}_{l})]\Psi_{a}^{\ast}\Psi_{a}.
\]

We apply the Lagrange multiplier method\cite{cou} to find the evolution
equation of $\Psi_{a}(t)$ driven by a velocity gradient specified by
Eq.(\ref{bc}). First of all we symmetrize Eq.(\ref{bc}) respect to all
particles. Secondly, we multiply the expression with $\lambda_{\mu_{\sigma}%
\nu_{\sigma}}^{\gamma(\sigma)}(t)$ and integrate it over the time
interval\cite{zos} $[t_{1},t_{2}]$:
\[
\int_{t_{1}}^{t_{2}}dt\sum_{\Psi_{b}}P^{\Phi_{b}}\int d\mathbf{R}%
f(\mathbf{r}_{\mu_{\sigma}\nu_{\sigma}}^{(\sigma)}-\mathbf{R})
\]%
\[
\lambda_{\mu_{\sigma}\nu_{\sigma}}^{\gamma(\sigma)}(t)\int d\tau\sum_{j=1}%
^{N}\delta(\mathbf{r}_{j}-\mathbf{R})[\frac{i\hbar}{2}(\Psi_{b}\frac
{\partial\Psi_{b}^{\ast}}{\partial r_{j\gamma}}-\Psi_{b}^{\ast}\frac
{\partial\Psi_{b}}{\partial r_{j\gamma}})
\]%
\begin{equation}
-mv_{\gamma}(\mathbf{r}_{\mu_{\sigma}\nu_{\sigma}}^{(\sigma)},t)\Psi_{b}%
\Psi_{b}^{\ast}]=0, \label{cst}%
\end{equation}
where the $\delta$ function is produced when we change $\int d\tau_{1}$ in
Eq.(\ref{bc}) to $\int d\tau$ in Eq.(\ref{cst}). The macroscopic velocity at
every point on each plate is a constraint. The overall Lagrangian density
including the constraint (\ref{cst}) is%
\begin{equation}
\mathcal{L}^{c}=\mathcal{L}+\sum_{\gamma\sigma\mu_{\sigma}\nu_{\sigma}}%
\lambda_{\mu_{\sigma}\nu_{\sigma}}^{\gamma(\sigma)}(t)\sum_{\Psi_{b}}%
P^{\Phi_{b}}\int d\mathbf{R} \label{la}%
\end{equation}%
\[
f(\mathbf{r}_{\mu_{\sigma}\nu_{\sigma}}^{(\sigma)}-\mathbf{R})\sum_{j=1}%
^{N}\delta(\mathbf{r}_{j}-\mathbf{R})
\]%
\[
\lbrack\frac{i\hbar}{2}(\Psi_{b}\frac{\partial\Psi_{b}^{\ast}}{\partial
r_{j\gamma}}-\Psi_{b}^{\ast}\frac{\partial\Psi_{b}}{\partial r_{j\gamma}%
})-mv_{\gamma}(\mathbf{r}_{\mu_{\sigma}\nu_{\sigma}}^{(\sigma)},t)\Psi_{b}%
\Psi_{b}^{\ast}]
\]
where the Lagrange multiplier $\lambda_{\mu_{\sigma}\nu_{\sigma}}%
^{\gamma(\sigma)}(t)$ is a real function of time\cite{zos} with dimension
length$^{4}\cdot$time$^{-1}$. Roughly speaking, $\lambda_{\mu_{\sigma}%
\nu_{\sigma}}^{\gamma(\sigma)}(t)$ is a product of the $\gamma^{th}$ component
of velocity at point $\mathbf{r}_{\mu_{\sigma}\nu_{\sigma}}^{(\sigma)}$ of the
$\sigma^{th}$ plate and the disturbed volume of fluid by a molecule at
$\mathbf{r}_{\mu_{\sigma}\nu_{\sigma}}^{(\sigma)}$. We will see this in more
detail in Sec.\ref{cz}.

The equation of motion for $\Psi_{a}(t)$ can be obtained from the Euler
equation:%
\begin{equation}
\frac{\partial\mathcal{L}^{c}}{\partial\Psi_{a}^{\ast}}-\sum_{j=1}^{N}%
\nabla_{\mathbf{r}_{j}}\frac{\partial\mathcal{L}^{c}}{\partial\Psi
_{a\mathbf{r}_{j}}^{\ast}}-\frac{\partial}{\partial t}\frac{\partial
\mathcal{L}^{c}}{\partial\Psi_{at}^{\ast}}=0.\label{eu}%
\end{equation}
It is%
\begin{equation}
i\hbar\Psi_{at}=(H_{0}+H_{f}+H_{c}+H_{d})\Psi_{a},\label{eu1}%
\end{equation}
where%
\begin{equation}
H_{c}=P^{\Phi_{a}}\sum_{\gamma\sigma\mu_{\sigma}\nu_{\sigma}}\lambda
_{\mu_{\sigma}\nu_{\sigma}}^{\gamma(\sigma)}(t)\int d\mathbf{R}f(\mathbf{r}%
^{(\sigma)}-\mathbf{R})\label{eu3}%
\end{equation}%
\[
\times\sum_{j=1}^{N}\delta(\mathbf{r}_{j}-\mathbf{R})mv_{\gamma}%
(\mathbf{r}_{\mu_{\sigma}\nu_{\sigma}}^{(\sigma)},t),
\]
and%
\[
H_{d}=P^{\Phi_{a}}\sum_{\gamma\sigma\mu_{\sigma}\nu_{\sigma}}\lambda
_{\mu_{\sigma}\nu_{\sigma}}^{\gamma(\sigma)}(t)\int d\mathbf{R}f(\mathbf{r}%
_{\mu_{\sigma}\nu_{\sigma}}^{(\sigma)}-\mathbf{R})
\]%
\begin{equation}
\times\sum_{j=1}^{N}\{\delta(\mathbf{r}_{j}-\mathbf{R})i\hbar\frac{\partial
}{\partial r_{j\gamma}}+\frac{i\hbar}{2}\frac{\partial\delta(\mathbf{r}%
_{j}-\mathbf{R})}{\partial r_{j\gamma}}\}.\label{eu4}%
\end{equation}
One may notice that Eq.(\ref{eu1}) contains only $\Psi_{a}$ and its partial
derivatives. Other $\Psi_{b}$ do not appear. In this sense, the velocity
gradient is a special mechanical perturbation. On the other hand, the solution
$\Psi_{a}(t)$ of Eq.(\ref{eu1}) involves $M_{c}$ parameters $\{\lambda
_{\mu_{\sigma}\nu_{\sigma}}^{\gamma(\sigma)}(t)\}$, that should also satisfy
$M_{c}$ constraints (\ref{bc}) which mix all $\{\Psi_{b}(t)\}$. Thus velocity
gradient is also an internal disturbance\cite{kub,kyn}.

In $H_{c}$ and $H_{d}$, the integrals behind the summation signs are the
macroscopic momentum density and negative momentum density operator
respectively. Thus the subtracted momentum density automatically
happens\cite{pic} in $H_{c}+H_{d}$ , which has to be introduced in some
previous theories\cite{ori}.

Because $H_{d}\neq H_{d}^{\ast}$, one cannot find a system with wave function
$\Psi^{\prime}(\{\mathbf{r}_{j}\};t)=\Psi_{a}^{\ast}(\{\mathbf{r}_{j}\};-t)$
such that $i\hbar\partial\Psi^{\prime}/\partial t=(H_{0}+H_{f}+H_{c}%
+H_{d})\Psi^{\prime}$, so that the process described by Eq.(\ref{eu1}) is
irreversible\cite{tol}.

In classical mechanics, constraints can also be dealt with the Lagrange
multiplier method\cite{whi,zos,gant,pars}. A constraint produces a reaction
force on each particle in the system. The constraints (\ref{bc}) enforce the
interaction on the system from four the plates.

Defining the macroscopic velocity field of a fluid by Eq.(\ref{mfv}) has a
profound consequence: the resultant constraint (\ref{bc}) on wave function is
bilinear. The evolution equation (\ref{eu1}) for $\Psi_{a}(t)$ is linear and
does not include other states. The evolution of the state of system driven by
a velocity gradient is \textit{almost} determined in the framework of pure
mechanics. If we had defined the macroscopic velocity field of fluid as
\begin{equation}
v_{\gamma}(\mathbf{r},t)=\sum_{\Psi_{a}}P^{\Phi_{a}}\int d\mathbf{R}%
f(\mathbf{r}-\mathbf{R})v_{\gamma}^{\Psi_{a}}(\mathbf{R},t),\label{mav}%
\end{equation}
with%
\begin{equation}
v_{\gamma}^{\Psi_{a}}(\mathbf{R},t)=S_{\gamma}^{\Psi_{a}}(\mathbf{R}%
,t)/\rho^{\Psi_{a}}(\mathbf{R},t),\label{miv}%
\end{equation}
we would have two serious difficulties: (i) this definition breaks the basic
requirement that the \textit{macroscopic} mass flux density $S_{\gamma
}(\mathbf{r},t)$ must always be the \textit{macroscopic} momentum of unit
volume of fluid\cite{v6}; (ii) the resulted temporal evolution equation for
$\Psi_{a}(t)$ is nonlinear, and $\Psi_{a}(t)$ couples with other $\{\Psi
_{b}(t)\}$. Certainly, assuming the fluid is incompressible, the labyrinthine
evolution equations could be linearized and decoupled.

\subsection{A comparison of velocity gradient and a mechanical disturbance}

\label{bj}

For a mechanical perturbation, external fields are some specified functions of
time. The interaction Hamiltonian $H_{f}$ is expressed by the external fields,
the coordinate and momentum operators of all particles\cite{kyn,short,jp}. The
states $\{\Psi_{a}(t)\}$ of system are not involved in $H_{f}$. On the other
hand, the macroscopic velocity field of a fluid is defined in Eq.(\ref{mfv})
through macroscopic mass density (\ref{md}) and macroscopic mass flux
(\ref{ond}). The later two quantities involve the states $\{\Psi_{a}(t)\}$ of
system, spatial coarse-graining and average over various initial conditions.
Before specifying the velocity gradient, which involves the states $\{\Psi
_{a}(t)\}$ of system, we cannot determine the evolution of the state of system
driven by a velocity gradient.

From the charge conservation law in a pure state, one can \textit{directly}
use the time-dependent Schrodinger equation (\ref{tse}) to
derive\cite{short,pss} the current density $\mathbf{j}^{\Psi_{a}}%
(\mathbf{r},t)$ in state $\Psi_{a}$ (the microscopic response) from the
temporal change of charge density $\rho^{\Psi_{a}}(\mathbf{r},t)$ in state
$\Psi_{a}(t)$. One does not need the quantal operator of current
density\cite{pre}, although the results are the same from the two points of
view. In contrast, we cannot use mass conservation in a pure state to derive
mass flux $S_{\gamma}^{\Psi_{a}}(\mathbf{r},t)$ in state $\Psi_{a}(t)$. We
have to invoke the definition of momentum density (\ref{dc}). In this sense,
the velocity gradient is an internal disturbance\cite{kub,kyn}. The states
$\{\Psi_{a}(t)\}$ of system driven by velocity gradient have to be determined
by the equation of motion (\ref{eu1}) and the constraints (\ref{bc}) self-consistently.

\section{Conservation laws in a pure state}

\label{shou}

In this section, we assume that $\lambda_{\mu_{\sigma}\nu_{\sigma}}%
^{\gamma(\sigma)}(t)$ and $\{\Psi_{b}(t)\}$ are known. The time rate of change
of energy, mass density and momentum density are derived. These relations are
formal but exact. We will find $\lambda_{\mu_{\sigma}\nu_{\sigma}}%
^{\gamma(\sigma)}(t)$ and $\{\Psi_{b}(t)\}$ in Sec.\ref{dirac}.

\subsection{Time rate of change of total energy}

\label{pow}

The temporal or spatial change in \textit{local} energy \textit{density} is
closely related to the temperature gradient. Since $\mathcal{S}$ is a member
of the canonical ensemble at temperature $T$, we cannot talk about the time
rate of change of \textit{local} energy \textit{density}.

For an isolated system $\mathcal{S}$, the total energy is conserved:%
\begin{equation}
E^{\Phi_{a}}(t)=\int d\tau e^{itE_{a}/\hbar}\Phi_{a}^{\ast}H_{0}\Phi
_{a}e^{-itE_{a}/\hbar}=E_{a}. \label{ise}%
\end{equation}
Consider a more general state of isolated system $\mathcal{S}$ with initial
value $\Psi(0)=\sum_{e}c_{e}\Phi_{e}$, where $\{c_{e}\}$ are some constants.
The time evolution of $\Psi(0)$ is $\Psi(t)=\sum_{e}c_{e}\Phi_{e}%
e^{-iE_{e}t/\hbar}$. From the orthogonality of eigenfunctions of $H_{0}$, it
is easy to check that the average energy of $\Psi(t)$ is conserved:
\begin{equation}
E^{\Psi(0)}=\int d\tau\Psi^{\ast}(0)H_{0}\Psi(0)=\sum_{e}|c_{e}|^{2}E_{e},
\label{ie1}%
\end{equation}
and%
\begin{equation}
E^{\Psi(t)}=\int d\tau\Psi^{\ast}(t)H_{0}\Psi(t)=\sum_{e}|c_{e}|^{2}E_{e}.
\label{ie2}%
\end{equation}
\

Now assume that the system $\mathcal{S}$ is in good contact with a thermal
reservoir $\mathcal{B}$ in temperature $T$. Thus, even if we apply a velocity
gradient and an external field on the system, $\mathcal{S}$ is maintained at
the same temperature $T$. Because $\mathcal{S}$ is macroscopic, in a time
period much longer than a macroscopic measurement, $\mathcal{S}$ can be viewed
as quasi-closed\cite{v5}. Therefore we can use a wave function $\Psi_{a}(t)$
to describe the system. The average energy of $\mathcal{S}$ in state $\Psi
_{a}(t)$ is defined as%

\begin{equation}
E^{\Psi_{a}}(t)=\int d\tau\Psi_{a}^{\ast}(t)H_{0}\Psi_{a}(t).\label{ets}%
\end{equation}
Because $\Psi_{a}(t)$ satisfies Eq.(\ref{eu1}), one can get the time rate of
change $dE^{\Psi_{a}}(t)/dt$ of the energy $E^{\Psi_{a}}(t)$ in pure state
$\Psi_{a}(t)$. The average energy $E(t)$ of $\mathcal{S}$ at time $t$ is
$E(t)=\sum_{\Psi_{a}}P^{\Phi_{a}}E^{\Psi_{a}}(t)$. The time rate of energy
exchange $dE(t)/dt$ between $\mathcal{S}$ and $\mathcal{B}$ is determined by
\begin{equation}
\frac{dE(t)}{dt}=\int d\mathbf{R}V(\mathbf{R},t)\nabla_{\mathbf{R}}%
\cdot\label{tre}%
\end{equation}%
\[
\sum_{a}P^{\Phi_{a}}\frac{i\hbar}{2m}N\int d\tau_{1}(\Psi_{a}\nabla
_{\mathbf{R}}\Psi_{a}^{\ast}-\Psi_{a}^{\ast}\nabla_{\mathbf{R}}\Psi_{a})
\]%
\[
-\sum_{\gamma\sigma\mu_{\sigma}\nu_{\sigma}}\lambda_{\mu_{\sigma}\nu_{\sigma}%
}^{\gamma(\sigma)}\int d\mathbf{R}f(\mathbf{r}_{\mu_{\sigma}\nu_{\sigma}%
}^{(\sigma)}-\mathbf{R})\frac{i\hbar}{2}\frac{\partial}{\partial R_{\gamma}%
}\nabla_{\mathbf{R}}\cdot
\]%
\[
\sum_{a}(P^{\Phi_{a}})^{2}\frac{i\hbar}{2m}N\int d\tau_{1}(\Psi_{a}%
\nabla_{\mathbf{R}}\Psi_{a}^{\ast}-\Psi_{a}^{\ast}\nabla_{\mathbf{R}}\Psi_{a})
\]%
\[
+\sum_{\gamma\sigma\mu_{\sigma}\nu_{\sigma}}\lambda_{\mu_{\sigma}\nu_{\sigma}%
}^{\gamma(\sigma)}\int d\mathbf{R}f(\mathbf{r}_{\mu_{\sigma}\nu_{\sigma}%
}^{(\sigma)}-\mathbf{R})mv_{\gamma}(\mathbf{r}_{\mu_{\sigma}\nu_{\sigma}%
}^{(\sigma)},t)\nabla_{\mathbf{R}}\cdot
\]%
\[
\sum_{a}(P^{\Phi_{a}})^{2}\frac{i\hbar}{2m}N\int d\tau_{1}(\Psi_{a}%
\nabla_{\mathbf{R}}\Psi_{a}^{\ast}-\Psi_{a}^{\ast}\nabla_{\mathbf{R}}\Psi_{a})
\]%
\[
+\sum_{\gamma\sigma\mu_{\sigma}\nu_{\sigma}}\lambda_{\mu_{\sigma}\nu_{\sigma}%
}^{\gamma(\sigma)}\int d\mathbf{R}f(\mathbf{r}_{\mu_{\sigma}\nu_{\sigma}%
}^{(\sigma)}-\mathbf{R})N\frac{i\hbar}{2}\nabla_{\mathbf{R}}\cdot
\]%
\[
\sum_{a}(P^{\Phi_{a}})^{2}\frac{i\hbar}{2m}\int d\tau_{1}(\frac{\partial
\Psi_{a}}{\partial R_{\gamma}}\nabla_{\mathbf{R}}\Psi_{a}^{\ast}-\Psi
_{a}^{\ast}\nabla_{\mathbf{R}}\frac{\partial\Psi_{a}}{\partial R_{\gamma}})
\]%
\[
-\sum_{\gamma\sigma\mu_{\sigma}\nu_{\sigma}}\lambda_{\mu_{\sigma}\nu_{\sigma}%
}^{\gamma(\sigma)}\int d\mathbf{R}f(\mathbf{r}_{\mu_{\sigma}\nu_{\sigma}%
}^{(\sigma)}-\mathbf{R})N(N-1)
\]%
\[
\sum_{a}(P^{\Phi_{a}})^{2}\int d\tau_{1}U(\mathbf{R},\mathbf{r}_{2}%
)\frac{\partial(\Psi_{a}\Psi_{a}^{\ast})}{\partial r_{2\gamma}},
\]
where the arguments of $\Psi_{a}$ are$\ (\mathbf{R},\mathbf{r}_{2}%
,\cdots,\mathbf{r}_{N};t)$. The last term is the power of the internal force
near the boundary surfaces. In classical mechanics, for a group of particles,
the internal forces do work on the system\cite{qin}. Only for a rigid body, is
the work done by internal forces zero. It is interesting to notice that only
for the regions close to the boundary plates [$f(\mathbf{r}_{\mu_{\sigma}%
\nu_{\sigma}}^{(\sigma)}-\mathbf{R})$ decays very rapidly when $|\mathbf{r}%
_{\mu_{\sigma}\nu_{\sigma}}^{(\sigma)}-\mathbf{R}|>k_{0}^{-1}$], the internal
forces do work.

In Eq.(\ref{tre}), we can see an obvious distinction between a velocity
gradient and a mechanical perturbation. The first term contains $P^{\Phi_{a}}%
$, is the power due to the external field. The remaining terms contains
$[P^{\Phi_{a}}]^{2}$, the square of the occupation probability of admissible
initial states. They are the contributions to power due to the friction at
boundary surfaces. The first $P^{\Phi_{a}}$ comes from prescribing the
macroscopic boundary condition as constraints on the wave function of a pure
state. The interaction with plates are expressed by $H_{c}+H_{d}$ in the
`Hamiltonian' of system, cf. Eqs.(\ref{eu3},\ref{eu4}). Since $dE^{\Psi_{a}%
}(t)/dt$ is determined by $\Psi_{at}(t)$, from Eq.(\ref{eu}), $dE^{\Psi_{a}%
}(t)/dt$ includes a $P^{\Phi_{a}}$. To obtain $dE(t)/dt$ from $dE^{\Psi_{a}%
}(t)/dt$, the average about possible initial states (\ref{ea}) introduced the
second $P^{\Phi_{a}}$.

The temperature of $\mathcal{S}$ is maintained at $T$ by the bath
$\mathcal{B}$. All the energy generated in the system is transferred to
$\mathcal{B}$. The entropy $S^{\mathcal{B}}(t)$ of the bath $\mathcal{B}$
increases with time according to
\begin{equation}
dS^{\mathcal{B}}(t)/dt=T^{-1}dE(t)/dt. \label{entr}%
\end{equation}
Substituting Eq.(\ref{tre}) into Eq.(\ref{entr}) and noticing that
$\lambda_{\mu_{\sigma}\nu_{\sigma}}^{\gamma(\sigma)}(t)f(\mathbf{r}%
_{\mu_{\sigma}\nu_{\sigma}}^{(\sigma)}-\mathbf{R})$ has dimension of velocity,
each term is in the form of temperature$^{-1}$ $\times$velocity$\times
$momentum$\times$velocity gradient, except the first and last term. This is
consistent with the macroscopic law of entropy production rate\cite{v6}.

\subsection{Time rate of change of mass density}

\label{zu}

With the help of Eq.(\ref{eu1}), one may calculate the time rate of
change\cite{ir50,ori} of $\rho^{\Psi_{a}}(\mathbf{r},t)$. Further applying the
averaging procedure (\ref{cg},\ref{ea}) to the expression for $\partial
\rho^{\Psi_{a}}(\mathbf{r},t)/\partial t$, one obtains the macroscopic law of
mass conservation:%
\begin{equation}
\frac{\partial\rho(\mathbf{r},t)}{\partial t}+\sum_{\gamma}\frac{\partial
S_{\gamma}^{\prime}(\mathbf{r},t)}{\partial r_{\gamma}}=0,\label{hmc0}%
\end{equation}
where%
\begin{equation}
S_{\gamma}^{\prime}(\mathbf{r},t)=S_{\gamma}(\mathbf{r},t)-\int d\mathbf{R}%
^{\prime}f(\mathbf{r}-\mathbf{R}^{\prime})\sum_{\sigma\mu_{\sigma}\nu_{\sigma
}}\lambda_{\mu_{\sigma}\nu_{\sigma}}^{\gamma(\sigma)}\label{hmc}%
\end{equation}%
\[
\times f(\mathbf{r}_{\mu_{\sigma}\nu_{\sigma}}^{(\sigma)}-\mathbf{R}^{\prime
})\sum_{\Psi_{a}}[P^{\Phi_{a}}]^{2}\rho^{\Psi_{a}}(\mathbf{R}^{\prime},t).
\]
In the R.H.S of Eq.(\ref{hmc}), the first term $S_{\gamma}(\mathbf{r},t)$ is
the usual macroscopic mass flux, and is given by Eq.(\ref{ond}). The second
term is an additional mass flux caused by the given velocity gradient. Two
truncation functions\cite{mel2,bi} $f(\mathbf{r}_{\mu_{\sigma}\nu_{\sigma}%
}^{(\sigma)}-\mathbf{R}^{\prime})$ and $f(\mathbf{r}-\mathbf{R}^{\prime})$
indicate that the correction is nonzero only near the boundary surfaces.
Again, $[P^{\Phi_{a}}]^{2}$ reflects the fact that a velocity gradient is an
internal disturbance.

\subsection{Microscopic response: momentum flux in a pure state}

\label{mfu}


Taking the time derivative of Eq.(\ref{mod}), one obtains the time rate of
change of the momentum density\cite{ir50,ori} $S_{\alpha}^{\Psi_{a}%
}(\mathbf{r},t)$ in state $\Psi_{a}(t)$. In terms of Eq.(\ref{eu1}), one has:%
\begin{equation}
\frac{\partial S_{\alpha}^{\Psi_{a}}(\mathbf{r},t)}{\partial t}+\sum_{\beta
}\frac{\partial\Pi_{\alpha\beta}^{\Psi_{a}}(\mathbf{r},t)}{\partial r_{\beta}%
}=-\frac{\partial V(\mathbf{r},t)}{\partial r_{\alpha}}\rho^{\Psi_{a}%
}(\mathbf{r},t)/m \label{cond}%
\end{equation}%
\[
+P^{\Phi_{a}}\sum_{\gamma\sigma\mu_{\sigma}\nu_{\sigma}}\lambda_{\mu_{\sigma
}\nu_{\sigma}}^{\gamma(\sigma)}(t)\frac{\partial f(\mathbf{r}_{\mu_{\sigma}%
\nu_{\sigma}}^{(\sigma)}-\mathbf{r})}{\partial r_{\alpha}}%
\]%
\[
\times\lbrack S_{\gamma}^{\Psi_{a}}(\mathbf{r},t)-v_{\gamma}(\mathbf{r}%
_{\mu_{\sigma}\nu_{\sigma}}^{(\sigma)},t)\rho^{\Psi_{a}}(\mathbf{r},t)],
\]
where $\alpha,$ $\beta,$ $\gamma=x,y,z$, and the momentum flux $\Pi
_{\alpha\beta}^{\Psi_{a}}(\mathbf{r},t)$ in state $\Psi_{a}$ is%
\begin{equation}
\Pi_{\alpha\beta}^{\Psi_{a}}(\mathbf{r},t)=\frac{\hbar^{2}}{2m}N\int d\tau_{1}
\label{flu}%
\end{equation}%
\[
\lbrack\frac{\partial\Psi_{a}}{\partial r_{\alpha}}\frac{\partial\Psi
_{a}^{\ast}}{\partial r_{\beta}}+\frac{\partial\Psi_{a}^{\ast}}{\partial
r_{\alpha}}\frac{\partial\Psi_{a}}{\partial r_{\beta}}-\frac{1}{2}%
\frac{\partial^{2}(\Psi_{a}\Psi_{a}^{\ast})}{\partial r_{\beta}\partial
r_{\alpha}}]
\]%
\[
-\frac{1}{2}N\int d\tau_{1}\sum_{k=2}^{N}(\mathbf{r}_{\beta}-\mathbf{r}%
_{k\beta})\frac{\partial U(\mathbf{r}_{k},\mathbf{r})}{\partial r_{\alpha}%
}\Psi_{a}\Psi_{a}^{\ast}.
\]
In Eq.(\ref{flu}), the arguments of $\Psi_{a}$ are $(\mathbf{r},\mathbf{r}%
_{2}\cdots\mathbf{r}_{N};t)$. To obtain the divergence form for the second
term in Eq.(\ref{flu}), we used the fact that the range of interaction force
is much shorter than\cite{mel2,zub} $k_{0}^{-1}$.

Although velocity gradient is an internal disturbance, by viewing the velocity
gradient as a constraint on the wave function, one can still define a
microscopic response: the momentum flux $\Pi_{\alpha\beta}^{\Psi_{a}}$ in
state $\Psi_{a}(t)$. Applying Eqs.(\ref{cg},\ref{ea}), one obtains the
measured macroscopic momentum flux:%
\begin{equation}
\Pi_{\alpha\beta}(\mathbf{r},t)=\sum_{\Psi_{a}}P^{\Phi_{a}}\int d\mathbf{R}%
f(\mathbf{r}-\mathbf{R})\Pi_{\alpha\beta}^{\Psi_{a}}(\mathbf{R},t).\label{mdu}%
\end{equation}
After $\Psi_{a}(t)$ is obtained, the shear viscosity can be read out from
Eqs.(\ref{flu},\ref{mdu}).

In the RHS of Eq.(\ref{cond}), the first term is the momentum density produced
by the external field per unit time, the second term is the momentum density
produced by the constraint (\ref{bc}) per unit time. Because the moving plate
drives the fluid, $\mathcal{S}$ gains momentum and energy. Thus there are
source terms in the equations for the time rate of change, cf. Eqs.
(\ref{cond}) and (\ref{tre}). Since no mass is produced by a velocity
gradient, there is not a source term in the time rate of change of mass
density, cf. Eq.(\ref{hmc0}).

There is a subtle difference between Eqs.(\ref{cond},\ref{flu}) and the
corresponding results in previous
theories\cite{mac1,mac2,mel2,zub,rob,rob66,rob67,pic,ir50,ir51}. In all
previous theories, conservation laws and fluxes (responses) are derived for an
isolated fluid system without imposing velocity gradient. The system
satisfies
\begin{equation}
i\hbar\partial\Psi_{0}/\partial t=H_{0}\Psi_{0} \label{fs}%
\end{equation}
or Newton's equation for an isolated
system\cite{mel2,zub,rob,rob66,rob67,pic,ir50,ir51}. The operator of momentum
flux is\cite{ori,ir50,ir51,rob}%
\[
\widehat{\Pi}_{\alpha\beta}(\mathbf{r})=\sum_{i=1}^{N}\frac{1}{4m}\{p_{\alpha
}^{i}p_{\beta}^{i}\delta(\mathbf{r}_{i}-\mathbf{r})+p_{\alpha}^{i}%
\delta(\mathbf{r}_{i}-\mathbf{r})p_{\beta}^{i}%
\]%
\[
+p_{\beta}^{i}\delta(\mathbf{r}_{i}-\mathbf{r})p_{\alpha}^{i}+\delta
(\mathbf{r}_{i}-\mathbf{r})p_{\beta}^{i}p_{\alpha}^{i}\}
\]%
\begin{equation}
+\frac{1}{2}\sum_{i=1}^{N}\sum_{j(\neq i)}(r_{\beta}^{i}-r_{\beta}^{j}%
)[-\frac{\partial U(\mathbf{r}_{i},\mathbf{r}_{j})}{\partial r_{\alpha}^{i}%
}]\delta(\mathbf{r}_{i}-\mathbf{r}). \label{dfu}%
\end{equation}
To obtain the viscosity, one must average operator (\ref{dfu}) over a
non-equilibrium density matrix (or distribution function) $\rho_{non}$ which
accommodates the effects of velocity
gradient\cite{mel2,zub,rob,rob66,rob67,pic,ir50,ir51}. This reminds us of some
older theories for the response to an electromagnetic field. In those
theories, the current operator is derived from the Schrodinger equation for an
isolated system (without external electromagnetic field). The resulting
current density missed the vector potential term and broke gauge
invariance\cite{cb,short,pre}.

In Eq.(\ref{flu}), $\Psi_{a}(t)$ is the solution of Eq.(\ref{eu1}) rather than
that of Eq.(\ref{fs}). Thus the influence of velocity gradient has been
included in $\Pi_{\alpha\beta}^{\Psi_{a}}(\mathbf{r},t)$. If we know $\Psi
_{a}(t)$, Eq.(\ref{flu}) directly gives the microscopic response $\Pi
_{\alpha\beta}^{\Psi_{a}}(\mathbf{r},t)$ to a velocity gradient in state
$\Psi_{a}(t)$. We will compare the results derived from two method in
Sec.\ref{fph}. At this point, we only note that Eq.(\ref{flu}) reduces to the
ordinary conservation law in the bulk, and is also correct near to the
boundary surfaces. For the effects which are first order in velocity gradient,
Eq.(\ref{cond},\ref{flu}) gives the same results as those based on
Eqs.(\ref{fs},\ref{dfu}).

\section{spatial coarse-graining}

\label{put}

Let us check whether the average procedure (\ref{cg},\ref{ea}) is adequate and
enough to describe the dissipation produced by a velocity gradient.

\subsection{Spatial coarse-graining includes temporal coarse-graining}

\label{tica}

In a liquid, suppose that the force between two neighboring molecules is
$U/d$, where $d$ is the average distance between two molecules, $U$ is the
interaction energy between two molecules at a distance $d$. The time needed to
relax a change $\hbar/d$ in momentum is $t_{m1}\thicksim\hbar/U$. The time
needed for a molecule to move a distance $d$ is $t_{m2}\thicksim d/(\hbar
/md)$. Near the equilibrium state, the potential energy is balanced by kinetic
energy $U\thicksim\hbar^{2}/(md^{2})$, and the two microscopic relaxation
times are the same order $t_{m1}\thicksim t_{m2}$ and will be denoted as
$t_{m}$.

We show that the spatial coarse-grained average contains temporal
coarse-graining automatically. Applying the spatial coarse-graining average
(\ref{cg}) to the microscopic law of mass conservation in a pure state
$\Psi_{a}$, the time rate of change of a conserved quantity must be the same
order as the divergence of the corresponding flux $\partial\rho^{c\Psi_{a}%
}(\mathbf{r},t)/\partial t\thicksim\sum_{\gamma}\partial S_{\gamma}^{c\Psi
_{a}}(\mathbf{r},t)/\partial x_{\gamma}$, i.e.%
\begin{equation}
\frac{mn}{t}\thicksim k_{0}\frac{\hbar}{d}n,\label{mc1}%
\end{equation}
where $n\thicksim d^{-3}$ is the number density of fluid. The relaxation time
$t_{c1}$ for a fluid drop is $t_{c1}\thicksim md/(k_{0}\hbar)$. Because the
cut-off wave vector is $k_{0}$, the time needed for a molecule to pass a
distance $k_{0}^{-1}$ is $k_{0}^{-1}/(\hbar/md)\thicksim(k_{0}d)^{-1}t_{m2}$,
the same order as $t_{c1}$. The temporal coarse-graining is automatically
realized in the conservation law of mass by the spatial coarse-graining
(\ref{cg}).

Applying spatial coarse-graining to the microscopic law of momentum
conservation in a pure state $\Psi_{a}$, one has $\partial S_{\gamma}%
^{c\Psi_{a}}(\mathbf{r},t)/\partial t\thicksim\sum_{\beta}\partial\Pi
_{\gamma\beta}^{c\Psi_{a}}(\mathbf{r},t)/\partial r_{\beta}$, i.e.%
\[
\frac{n}{t}\frac{\hbar}{d}\thicksim k_{0}\frac{\hbar}{d}n\cdot\frac{\hbar}%
{md},
\]
the relaxation time of momentum in a fluid drop is the same as $t_{c1}$. The
temporal coarse-graining is automatically realized in the conservation law of
momentum by the spatial coarse-graining (\ref{cg}). We do not have energy relaxation.

We derive the coarse-grained time scale $t_{c1}$ from another point of view.
There are $N_{1}\thicksim(k_{0}d)^{-3}$ molecules in a coarse-grained fluid
drop. Because the range of molecular interaction is only extended to $d$, a
fluid drop interacts with its surrounding only through its surface. Thus the
force on a fluid drop is $N_{1}^{2/3}U/d$. The acceleration of the fluid drop
is $a\thicksim(N_{1}^{2/3}U/d)/(mN_{1})$, the time needed for the fluid drop
to move a distance $s=N_{1}^{1/3}d$ is $t_{c2}\thicksim(s/a)^{1/2}\thicksim
k_{0}^{-1}(m/U)^{1/2}\thicksim(k_{0}d)^{-1}(\hbar/U)$. Two characteristic
times for the coarse-grained fluid drop are the same order $t_{c1}\thicksim
t_{c2}$, will be denoted as $t_{c}$. Because $k_{0}^{-1}\gg d$, we can see
$t_{m}\ll t_{c}$, the coarse-grained time scale is much longer than the
microscopic time scale $t_{c2}\thicksim(k_{0}d)^{-1}t_{m}$.

The time coarse graining induced by the spatial coarse graining is fine enough
to describe\ macroscopic motion. The shear viscosity is order of
$\eta\thicksim mU/(\hbar d)$. In a macroscopic measurement, the acceleration
on a fluid drop produced by the velocity gradient is $a_{M}\thicksim\eta
\frac{dv_{x}}{dy}k_{0}^{-2}[mnk_{0}^{-3}]^{-1}$. The macroscopic relaxation
time is order of $t_{M}=v_{x0}/a_{M}\thicksim\lbrack v_{x0}(\frac{dv_{x}}%
{dy}d)^{-1}]t_{c}\thicksim t_{c}l_{y}/d$, where $l_{y}$ is a macroscopic
length. Therefore $t_{M}\gg t_{c}$. In summary, we have%
\begin{equation}
t_{m}\ll t_{c}\ll t_{M}. \label{3}%
\end{equation}

For the thermal motion of molecule, we could make a similar estimation: the
velocity and kinetic energy of a molecule should use $v_{T}=(k_{B}T/m)^{1/2}$
and $k_{B}T$ respectively. Relation (\ref{3}) holds. In conclusion, the
spatial coarse-graining (\ref{cg}) included a temporal coarse-graining, which
is adequate and enough to describe the dissipation produced by a velocity gradient.

\subsection{Spatial coarse-graining includes coarse-graining in the eigenvalue
spectrum of the collective variables}

\label{egcr}

In Sec.\ref{msm}, the quantal mean value of any operator in a pure state is
spatially coarse-grained with Eq.(\ref{cg}). Many authors choose to
coarse-grain the eigenvalue spectrum of collective variables. We will
illustrate that the spatial coarse-graining in the microscopic response
includes coarse-graining in the eigenvalue spectrum of collective variables.

Because the cut-off wave vector is $k_{0}$, there are $N_{1}\thicksim
(k_{0}d)^{-3}$ molecules in a coarse-grained fluid drop. A collective variable
for a liquid drop is a symmetric sum of single-particle variables\cite{if}.
The typical interval $\Delta_{N_{1}}$ between two eigenvalues of a collective
variable is $\Delta_{N_{1}}\thicksim\Delta_{1}e^{-N_{1}}$, where $\Delta_{1}$
is the typical interval between two eigenvalues of the corresponding
single-particle variable\cite{v5}. Excepting macroscopic quantum phenomenon
(superconductivity and superfluidity), even in a microscopic relaxation time
$t_{m2}\thicksim md^{2}/\hbar$, the action for a drop is $N_{1}(\hbar
^{2}/md^{2})(md^{2}/\hbar)\thicksim(k_{0}d)^{-3}\hbar\gg\hbar$. Therefore the
collective variables are classical variables. Since the relative error for a
collective variable is\cite{v5} $\Delta f/f\thicksim N_{1}^{-1/2}$, the
measurement error for the energy of a fluid drop is $\Delta_{N_{1}}\thicksim
N_{1}^{1/2}(\hbar^{2}/md^{2})$. The energy interval for a single molecules is
$\Delta_{1}\thicksim\hbar^{2}/md^{2}$. A macroscopic energy measurement with
spatial resolution $k_{0}^{-1}$ must involve $N_{1}^{1/2}e^{N_{1}}$
eigenvalues and corresponding eigenvectors of the energy operator of a liquid drop.

\section{calculating Lagrange multipliers}

\label{dirac}

In linear response theory, to obtain the observed macroscopic response to a
time-dependent mechanical perturbation, Dirac's method of the variation of
constants is frequently used to solve the Liouville equation\cite{k57} or
Schrodinger equation\cite{deg,short,pss,4t,epl}. We first use Dirac's
perturbation theory to get Lagrange multipliers, and then discuss its
applicable conditions.

\subsection{Lagrange multiplier}

\label{cz}

Consider a state $\Psi_{a}(t)$ with initial condition in the remote past
$\Psi_{a}(t=-\infty)=\Phi_{a}$. In first order of $H_{1}(t)=H_{f}%
(t)+H_{c}(t)+H_{d}$, the normalized wave function is%
\begin{equation}
\Psi_{a}(t)=e^{-iE_{a}t/\hbar}\Phi_{a}(1+i\operatorname{Im}g_{aa}^{(1)}%
)+\sum_{j(\neq a)}g_{ja}^{(1)}(t)e^{-iE_{j}t/\hbar}\Phi_{j}. \label{sup}%
\end{equation}
For both $j=a$ and $j\neq a$, the first order expansion coefficient is given
by:%
\begin{equation}
g_{ja}^{(1)}(t)=c_{ja}^{f}(t)+\sum_{\gamma\sigma\mu_{\sigma}\nu_{\sigma}%
}\lambda_{\mu_{\sigma}\nu_{\sigma}}^{\gamma(\sigma)}(t)[Z_{\mu_{\sigma}%
\nu_{\sigma}}^{\gamma(\sigma)}]_{ja}(t). \label{1c}%
\end{equation}
The second term in Eq.(\ref{1c}) is the contribution from the moving plate and
three static surfaces:%
\[
\lbrack Z_{\mu_{\sigma}\nu_{\sigma}}^{\gamma(\sigma)}]_{cb}(t)=\frac{1}%
{2}P^{\Phi_{b}}\int_{-\infty}^{t}dt^{\prime}e^{i\omega_{cb}t^{\prime}}\int
d\mathbf{R}f(\mathbf{r}_{\mu_{\sigma}\nu_{\sigma}}^{(\sigma)}-\mathbf{R})
\]%
\[
\times N\int d\tau_{1}(\Phi_{c}^{\ast}\frac{\partial\Phi_{b}}{\partial
R_{\gamma}}-\Phi_{b}\frac{\partial\Phi_{c}^{\ast}}{\partial R_{\gamma}})
\]%
\[
-P^{\Phi_{b}}\frac{i}{\hbar}mv_{\gamma}(\mathbf{r}_{\mu_{\sigma}\nu_{\sigma}%
}^{(\sigma)},t)\int_{-\infty}^{t}dt^{\prime}e^{i\omega_{ja}t^{\prime}}%
\]%
\begin{equation}
\times\int d\mathbf{R}f(\mathbf{r}_{\mu_{\sigma}\nu_{\sigma}}^{(\sigma
)}-\mathbf{R})N\int d\tau_{1}\Phi_{c}^{\ast}\Phi_{b}, \label{z1}%
\end{equation}
where $\omega_{ja}=(E_{j}-E_{a})/\hbar$. Denote $d$ as the average distance
between two neighboring molecules, $\omega$ the frequency of a monochromatic
driving velocity $v_{x}(\mathbf{r}^{(\sigma)},t)$, $[Z_{\mu_{\sigma}%
\nu_{\sigma}}^{\gamma(\sigma)}]_{ja}\thicksim\lbrack(\omega_{ja}+\omega
)d^{4}]^{-1}$ characterizes the transition caused by a point $\mathbf{r}%
_{\mu_{\sigma}\nu_{\sigma}}^{(\sigma)}$ at the $\sigma^{th}$ plate, and has
the dimension of time$\cdot$length$^{-4}$. The first term in Eq.(\ref{1c}) is
the contribution from the external field:%
\begin{equation}
c_{ja}^{f}(t)=\int_{-\infty}^{t}dt^{\prime}e^{i\omega_{ja}t^{\prime}}\int
d\tau\Phi_{j}^{\ast}[-\frac{i}{\hbar}\sum_{k=1}^{N}V(\mathbf{r}_{k},t^{\prime
})]\Phi_{a}. \label{fc}%
\end{equation}
The first term in the RHS of Eq.(\ref{sup}) is the free evolution term,
$i\operatorname{Im}g_{aa}^{(1)}(t)$ is caused by the requirement that $1=\int
d\tau\Psi_{a}^{\ast}(t)\Psi_{a}(t)$ should be correct to first order in
$H_{1}(t)$. In the kinetic theories, one concerns various transition
probabilities. We only need the $j\neq a$ terms in Eq.(\ref{sup}).

Before we apply the external field and move the top plate, the
system\ $\mathcal{S}$ is in equilibrium at the bath temperature $T$: the
system is in various stationary states \{$\Phi_{b}$\} according to
Eq.(\ref{canp}). In an ordinary fluid, there is no macroscopic mass flux at
any boundary surface $S_{\gamma}^{(0)}(\mathbf{r}_{\mu_{\sigma}\nu_{\sigma}%
}^{(\sigma)},t)=0$:%
\begin{equation}
\frac{i\hbar}{2}\sum_{b}P^{\Phi_{b}}\int d\mathbf{R}f(\mathbf{r}_{\mu_{\sigma
}\nu_{\sigma}}^{(\sigma)}-\mathbf{R})N\label{sta}%
\end{equation}%
\[
\times\int d\tau_{1}(\Phi_{b}\frac{\partial\Phi_{b}^{\ast}}{\partial
R_{\gamma}}-\Phi_{b}^{\ast}\frac{\partial\Phi_{b}}{\partial R_{\gamma}})=0,
\]
where $\gamma=x,y,z$, $\sigma=1,2,3,4$, the arguments of $\Phi_{b}$ are
($\mathbf{R},\mathbf{r}_{2},\cdots,\mathbf{r}_{N}$). Superfluidity is excluded
by the requirement (\ref{sta}).

To conserve space in the subsequent calculations, we rewrite Eq.(\ref{sup}) as%

\begin{equation}
\Psi_{a}(t)=e^{-iE_{a}t/\hbar}\Phi_{a}+\sum_{j}c_{ja}(t)e^{-iE_{j}t/\hbar}%
\Phi_{j}, \label{sup1}%
\end{equation}
where $c_{ja}(t)=i\operatorname{Im}g_{aa}^{(1)}(t)$ for $j=a$; $c_{ja}%
(t)=g_{ja}^{(1)}(t)$ for $j\neq a$. The summation includes both $j=a$ and
$j\neq a$. Substitute Eq.(\ref{sup1}) into boundary conditions (\ref{bc}), and
notice Eq.(\ref{sta}) for four boundary surfaces, then we have
\begin{equation}
v_{\gamma}(\mathbf{r}_{\mu_{\sigma}\nu_{\sigma}}^{(\sigma)},t)\rho^{0}%
=\sum_{bd}P^{\Phi_{b}}\int d\mathbf{R}f(\mathbf{r}_{\mu_{\sigma}\nu_{\sigma}%
}^{(\sigma)}-\mathbf{R})N\int d\tau_{1} \label{cst1}%
\end{equation}%
\[
\{\frac{i\hbar}{2}[c_{db}(t)e^{-i(E_{d}-E_{b})t/\hbar}(\Phi_{d}\frac
{\partial\Phi_{b}^{\ast}}{\partial R_{\gamma}}-\frac{\partial\Phi_{d}%
}{\partial R_{\gamma}}\Phi_{b}^{\ast})
\]%
\[
-c_{db}^{\ast}(t)e^{i(E_{d}-E_{b})t/\hbar}(\Phi_{d}^{\ast}\frac{\partial
\Phi_{b}}{\partial R_{\gamma}}-\frac{\partial\Phi_{d}^{\ast}}{\partial
R_{\gamma}}\Phi_{b})]
\]%
\[
-v_{\gamma}(\mathbf{r}_{\mu_{\sigma}\nu_{\sigma}}^{(\sigma)},t)m[c_{db}%
(t)e^{-i(E_{d}-E_{b})t/\hbar}\Phi_{d}\Phi_{b}^{\ast}%
\]%
\[
+c_{db}^{\ast}(t)e^{i(E_{d}-E_{b})t/\hbar}\Phi_{b}\Phi_{d}^{\ast}]\},
\]
where
\begin{equation}
\rho^{0}=m\int d\mathbf{R}f(\mathbf{r}_{\mu_{\sigma}\nu_{\sigma}}^{(\sigma
)}-\mathbf{R})\sum_{b}P^{\Phi_{b}}N\int d\tau_{1}\Phi_{b}\Phi_{b}^{\ast}
\label{cden}%
\end{equation}
is the constant mass density of liquid in absent of external field and
boundary conditions (\ref{bc}).

Making use of Eq.(\ref{1c}), Eq.(\ref{cst1}) is reduced to a group of linear
equations for Lagrange multipliers:%
\begin{equation}
\sum_{\gamma^{\prime}\sigma^{\prime}\mu_{\sigma^{\prime}}^{\prime}\nu
_{\sigma^{\prime}}^{\prime}}A[_{\mu_{\sigma}\nu_{\sigma}}^{\gamma(\sigma
)},_{\mu_{\sigma^{\prime}}^{\prime}\nu_{\sigma^{\prime}}^{\prime}}%
^{\gamma^{\prime}(\sigma^{\prime})}](t)\lambda_{\mu_{\sigma^{\prime}}^{\prime
}\nu_{\sigma^{\prime}}^{\prime}}^{\gamma^{\prime}(\sigma^{\prime})}%
(t)=B_{\mu_{\sigma}\nu_{\sigma}}^{\gamma(\sigma)}(t),\label{Li}%
\end{equation}
where%
\begin{equation}
B_{\mu_{\sigma}\nu_{\sigma}}^{\gamma(\sigma)}(t)=v_{\gamma}(\mathbf{r}%
_{\mu_{\sigma}\nu_{\sigma}}^{(\sigma)},t)\rho^{0}\label{L1}%
\end{equation}%
\[
-\int d\mathbf{R}f(\mathbf{r}_{\mu_{\sigma}\nu_{\sigma}}^{(\sigma)}%
-\mathbf{R})\sum_{bd(b\neq d)}P^{\Phi_{b}}N\int d\tau_{1}%
\]%
\[
\{c_{db}^{f}(t)e^{-it\omega_{db}}[\frac{i\hbar}{2}(\Phi_{d}\frac{\partial
\Phi_{b}^{\ast}}{\partial R_{\gamma}}-\frac{\partial\Phi_{d}}{\partial
R_{\gamma}}\Phi_{b}^{\ast})-v_{\gamma}(\mathbf{r}_{\mu_{\sigma}\nu_{\sigma}%
}^{(\sigma)},t)m\Phi_{d}\Phi_{b}^{\ast}]
\]%
\[
-c_{db}^{f\ast}(t)e^{it\omega_{db}}[\frac{i\hbar}{2}(\Phi_{d}^{\ast}%
\frac{\partial\Phi_{b}}{\partial R_{\gamma}}-\frac{\partial\Phi_{d}^{\ast}%
}{\partial R_{\gamma}}\Phi_{b})+v_{\gamma}(\mathbf{r}_{\mu_{\sigma}\nu
_{\sigma}}^{(\sigma)},t)m\Phi_{b}\Phi_{d}^{\ast}]\},
\]
and%
\begin{equation}
A[_{\mu_{\sigma}\nu_{\sigma}}^{\gamma(\sigma)},_{\mu_{\sigma^{\prime}}%
^{\prime}\nu_{\sigma^{\prime}}^{\prime}}^{\gamma^{\prime}(\sigma^{\prime}%
)}](t)=\int d\mathbf{R}f(\mathbf{r}_{\mu_{\sigma}\nu_{\sigma}}^{(\sigma
)}-\mathbf{R})\label{L2}%
\end{equation}%
\[
\sum_{bd(b\neq d)}P^{\Phi_{b}}N\int d\tau_{1}\{[Z_{\mu_{\sigma^{\prime}%
}^{\prime}\nu_{\sigma^{\prime}}^{\prime}}^{\gamma^{\prime}(\sigma^{\prime}%
)}]_{db}(t)e^{-it\omega_{db}}%
\]%
\[
\lbrack\frac{i\hbar}{2}(\Phi_{d}\frac{\partial\Phi_{b}^{\ast}}{\partial
R_{\gamma}}-\frac{\partial\Phi_{d}}{\partial R_{\gamma}}\Phi_{b}^{\ast
})-v_{\gamma}(\mathbf{r}_{\mu_{\sigma}\nu_{\sigma}}^{(\sigma)},t)m\Phi_{d}%
\Phi_{b}^{\ast}]
\]%
\[
-[Z_{\mu_{\sigma^{\prime}}^{\prime}\nu_{\sigma^{\prime}}^{\prime}}%
^{\gamma^{\prime}(\sigma^{\prime})}]_{db}^{\ast}(t)e^{it\omega_{db}}%
\]%
\[
\times\lbrack\frac{i\hbar}{2}(\Phi_{d}^{\ast}\frac{\partial\Phi_{b}}{\partial
R_{\gamma}}-\frac{\partial\Phi_{d}^{\ast}}{\partial R_{\gamma}}\Phi
_{b})+v_{\gamma}(\mathbf{r}_{\mu_{\sigma}\nu_{\sigma}}^{(\sigma)},t)m\Phi
_{b}\Phi_{d}^{\ast}]\}.
\]
The dimension of $B_{\mu_{\sigma}\nu_{\sigma}}^{\gamma(\sigma)}$ is momentum
density. The first term in Eq.(\ref{L1}) is the momentum density produced by a
point $\mathbf{r}^{(\sigma)}$ at the $\sigma^{th}$ plate $\thicksim v_{\gamma
}(\mathbf{r}^{(\sigma)},t)md^{-3}$. The second term is the momentum density
produced by the interference between the external field and the constraint of
velocity gradient $\thicksim\lbrack(\omega_{db}+\omega_{f})^{-1}%
(V_{db}/d)]d^{-3}$, where $\omega_{f}$ is the frequency of the external field.
$B_{\sigma}$ is the net momentum density at $\mathbf{r}^{(\sigma)}$ on the
$\sigma^{th}$ plate. The dimension of $\{A_{\sigma\sigma^{\prime}}\}$ is
momentum density$\cdot$time$\cdot$length$^{-4}$. $A[_{\mu_{\sigma}\nu_{\sigma
}}^{\gamma(\sigma)},_{\mu_{\sigma^{\prime}}^{\prime}\nu_{\sigma^{\prime}%
}^{\prime}}^{\gamma^{\prime}(\sigma^{\prime})}]\thicksim\lbrack(\omega
_{db}+\omega)d^{4}]^{-1}(\hbar/d)d^{-3}$ describes the effect of a point
$\mathbf{r}_{\mu_{\sigma^{\prime}}^{\prime}\nu_{\sigma^{\prime}}^{\prime}%
}^{(\sigma^{\prime})}$ at the $\sigma^{\prime}$th plate to the point
$\mathbf{r}_{\mu_{\sigma}\nu_{\sigma}}^{(\sigma)}$ on the $\sigma^{th}$ plate.
Eq.(\ref{Li}) describes the momentum density at point $\mathbf{r}_{\mu
_{\sigma}\nu_{\sigma}}^{(\sigma)}$ produced by all points on the four boundary
plates. The Lagrange multipliers $\lambda_{\mu_{\sigma}\nu_{\sigma}}%
^{\gamma(\sigma)}$ are determined by the coupled linear equations (\ref{Li}),
and they are in general nonlinear functions of the velocity gradient.

The lowest order approximation to $\lambda_{\mu_{\sigma}\nu_{\sigma}}%
^{\gamma(\sigma)}$ is first order in $v_{\gamma}(\mathbf{r}_{\mu_{\sigma}%
\nu_{\sigma}}^{(\sigma)},t)/l_{y}$:
\begin{equation}
\lambda_{\mu_{\sigma}\nu_{\sigma}}^{\gamma(\sigma)}\thicksim\lbrack v_{\gamma
}(\mathbf{r}_{\mu_{\sigma}\nu_{\sigma}}^{(\sigma)},t)/(\hbar/md)][(\omega
_{db}+\omega)d]d^{3}. \label{mda}%
\end{equation}
From Eq.(\ref{tre}), we can see that the lowest order dissipation rate is
second order in velocity gradient, which is consistent with the result derived
from Navier-Stokes equation and the second law of thermodynamics\cite{v6}. In
addition, from Eqs.(\ref{mda},\ref{eu3},\ref{eu4}), $H_{c}$ is second order
and $H_{d}$ is first order in velocity gradient. This is similar to the
coupling between electromagnetic field and material: there are both first
order and second order terms in the vector potential. The small parameter for
the perturbation caused by $H_{f}$ is $\langle b|V(\mathbf{r}_{1}%
,t)|c\rangle/(\hbar^{2}/md^{2})$, for the perturbation caused by $H_{d}$ is
$v_{\gamma}(\mathbf{r}_{\mu_{\sigma}\nu_{\sigma}}^{(\sigma)},t)/(\hbar/md)$,
for the perturbation caused by $H_{c}$ is $[v_{\gamma}(\mathbf{r}_{\mu
_{\sigma}\nu_{\sigma}}^{(\sigma)},t)/(\hbar/md)]^{2}$.

If we substitute the estimation (\ref{mda}) into Eq.(\ref{eu4}), one has%
\begin{equation}
H_{d}\thicksim\sum_{j}\{mv_{\gamma}(\mathbf{r}_{\mu_{\sigma}\nu_{\sigma}%
}^{(\sigma)},t)(\omega_{db}+\omega)d\}. \label{oneq}%
\end{equation}
The quantity in \{\} is the dissipative part of the phenomenological
non-equilibrium Hamiltonian for the $j^{th}$ molecule\cite{hoo,ean,eva}. Here
$(\omega_{db}+\omega)$ corresponds to the strain rate ($\omega$ the frequency
of a monochromatic driving), $d$ the displacement of an atom, $mv_{\gamma
}(\mathbf{r}_{\mu_{\sigma}\nu_{\sigma}}^{(\sigma)},t)$ the momentum of an atom
at point $\mathbf{r}_{\mu_{\sigma}\nu_{\sigma}}^{(\sigma)}$. Of course,
Eq.(\ref{oneq}) has not changed from the boundary surfaces to the
bulk\cite{eva,hoo,ean}.

From $\{\lambda_{\mu_{\sigma}\nu_{\sigma}}^{\gamma(\sigma)}(t)\}$,
Eqs.(\ref{z1},\ref{fc},\ref{1c}) tells us all the expansion coefficients in
Eq.(\ref{sup1}) i.e. the state $\Psi_{a}(t)$.

\subsection{Shear viscosity}

\label{nx}

Next, we notice that no momentum flux exists in an equilibrium state:%
\begin{equation}
\Pi_{\gamma\beta}^{(0)}(\mathbf{r},t)=\sum_{\Phi_{b}}P^{\Phi_{b}}\int
d\mathbf{R}f(\mathbf{r}-\mathbf{R})\Pi_{\gamma\beta}^{\Phi_{b}}(\mathbf{R}%
,t)=0, \label{0f}%
\end{equation}
where%
\begin{equation}
\Pi_{\gamma\beta}^{\Phi_{b}}(\mathbf{R},t)=\frac{\hbar^{2}}{2m}N\int d\tau_{1}
\label{0mf}%
\end{equation}%
\[
\times\lbrack\frac{\partial\Phi_{b}}{\partial R_{\gamma}}\frac{\partial
\Phi_{b}^{\ast}}{\partial R_{\beta}}+\frac{\partial\Phi_{b}^{\ast}}{\partial
R_{\gamma}}\frac{\partial\Phi_{b}}{\partial R_{\beta}}-\frac{1}{2}%
\frac{\partial^{2}(\Phi_{b}\Phi_{b}^{\ast})}{\partial R_{\beta}\partial
R_{\gamma}}]
\]%
\[
-\frac{1}{2}N\int d\tau_{1}\sum_{k=2}^{N}(R_{\beta}-\mathbf{r}_{k\beta}%
)\frac{\partial U(\mathbf{r}_{k},\mathbf{R})}{\partial R_{\gamma}}\Phi_{b}%
\Phi_{b}^{\ast},
\]
is the momentum flux in pure state $\Phi_{b}$.

By means of Eq.(\ref{sup1}), the momentum flux (\ref{flu}) in a pure state is
reduced to%

\begin{equation}
\Pi_{\gamma\beta}^{\Psi_{b}}(\mathbf{R},t)=\frac{\hbar^{2}}{2m}N\int d\tau
_{1}\{\sum_{d}c_{db}(t)e^{-it\omega_{db}}\label{fl}%
\end{equation}%
\[
\times(\frac{\partial\Phi_{b}^{\ast}}{\partial R_{\gamma}}\frac{\partial
\Phi_{d}}{\partial R_{\beta}}+\frac{\partial\Phi_{d}}{\partial R_{\gamma}%
}\frac{\partial\Phi_{b}^{\ast}}{\partial R_{\beta}}-\frac{1}{2}\frac
{\partial^{2}(\Phi_{d}\Phi_{b}^{\ast})}{\partial R_{\beta}\partial R_{\gamma}%
})
\]%
\[
+\sum_{d}c_{db}^{\ast}(t)e^{it\omega_{db}}(\frac{\partial\Phi_{b}}{\partial
R_{\gamma}}\frac{\partial\Phi_{d}^{\ast}}{\partial R_{\beta}}+\frac
{\partial\Phi_{d}^{\ast}}{\partial R_{\gamma}}\frac{\partial\Phi_{b}}{\partial
R_{\beta}}-\frac{1}{2}\frac{\partial^{2}(\Phi_{b}\Phi_{d}^{\ast})}{\partial
R_{\beta}\partial R_{\gamma}})\}
\]%
\[
-\frac{1}{2}N\int d\tau_{1}\sum_{k=2}^{N}(R_{\beta}-\mathbf{r}_{k\beta}%
)\frac{\partial U(\mathbf{r}_{k},\mathbf{R})}{\partial R_{\gamma}}%
\]%
\[
\times\{\sum_{d}c_{db}(t)e^{-it\omega_{db}}\Phi_{d}\Phi_{b}^{\ast}+\sum
_{d}c_{db}^{\ast}(t)e^{it\omega_{db}}\Phi_{b}\Phi_{d}^{\ast}\}.
\]
Applying the average procedure (\ref{cg},\ref{ea}) to Eq.(\ref{fl}), the
observed macroscopic momentum flux (\ref{mdu}) becomes
\begin{equation}
\Pi_{\alpha\beta}(\mathbf{r},t)=\int d\mathbf{R}f(\mathbf{r}-\mathbf{R}%
)\sum_{b}P^{\Phi_{b}}\frac{\hbar^{2}}{2m}N\int d\tau_{1}\label{mrf}%
\end{equation}%
\[
\{\sum_{d}c_{db}(t)e^{-it\omega_{db}}(\frac{\partial\Phi_{b}^{\ast}}{\partial
R_{\alpha}}\frac{\partial\Phi_{d}}{\partial R_{\beta}}+\frac{\partial\Phi_{d}%
}{\partial R_{\alpha}}\frac{\partial\Phi_{b}^{\ast}}{\partial R_{\beta}}%
-\frac{1}{2}\frac{\partial^{2}(\Phi_{d}\Phi_{b}^{\ast})}{\partial R_{\beta
}\partial R_{\alpha}})
\]%
\[
+\sum_{d}c_{db}^{\ast}(t)e^{it\omega_{db}}(\frac{\partial\Phi_{b}}{\partial
R_{\alpha}}\frac{\partial\Phi_{d}^{\ast}}{\partial R_{\beta}}+\frac
{\partial\Phi_{d}^{\ast}}{\partial R_{\alpha}}\frac{\partial\Phi_{b}}{\partial
R_{\beta}}-\frac{1}{2}\frac{\partial^{2}(\Phi_{b}\Phi_{d}^{\ast})}{\partial
R_{\beta}\partial R_{\alpha}})\}
\]%
\[
-\int d\mathbf{R}f(\mathbf{r}-\mathbf{R})\sum_{b}P^{\Phi_{b}}\frac{1}{2}N\int
d\tau_{1}\sum_{k=2}^{N}(R_{\beta}-\mathbf{r}_{k\beta})\frac{\partial
U(\mathbf{r}_{k},\mathbf{R})}{\partial R_{\alpha}}%
\]%
\[
\times\{\sum_{d}c_{db}(t)e^{-it\omega_{db}}\Phi_{d}\Phi_{b}^{\ast}+\sum
_{d}c_{db}^{\ast}(t)e^{it\omega_{db}}\Phi_{b}\Phi_{d}^{\ast}\}.
\]
Eq.(\ref{mrf}) is first order in $H_{1}(t)$. The order $v_{x0}/l_{y}$
contribution comes from $H_{d}$. To recover usual viscosity (the proportional
constant between momentum flux and velocity gradient $v_{x0}/l_{y}$), in
$c_{db}(t)$, we only need the first term of Eq.(\ref{z1}).

\section{Time scales: justification of the present scheme}

\label{tcd}

\subsection{Justification of the method of variation of constants}

Applying the method of variation of constants (MVC) to a macroscopic system
worries many scientists\cite{ka1,cb,wang,fan,vho,ka2,ka3}. The kinetic
approach insists that the long time behavior of a large system should be
described by a time coarse-grained master equation\cite{ka1,cb,vho,ka2,ka3}.
We will show that with some caution, the MVC is applicable to calculate shear viscosity.

\subsubsection{Wave function description}

We first estimate the time period during which the system can be described by
a wave function. The thermal reservoir plays the role of damping for
$\mathcal{S}$. Eq.(\ref{eu1}) should be revised as%
\begin{equation}
i\hbar\frac{\partial\Psi_{a}}{\partial t}=(H_{0}+H_{f}+H_{c}+H_{d})\Psi
_{a}-iPt\Psi_{a}, \label{os}%
\end{equation}
where $P$ is the net power exerted on $\mathcal{S}$. Usually the last term is
not written out, but is implicitly taken into account\cite{fan}. $\mathcal{S}$
receives the power generated by the external field and velocity gradient, in
the same time $\mathcal{S}$ delivers extra energy to $\mathcal{B}$ to maintain
the equilibrium with $\mathcal{B}$. If at the initial moment $t=-\infty$, the
system is exactly in a pure state $\Psi_{a}(-\infty)=\Phi_{a}$, $\Psi_{a}(t)$
is uniquely determined by Eq.(\ref{os}) and is finite for any $t>-\infty$.
Then Eq.(\ref{cond}) and the corresponding macroscopic equation obtained by
the averaging procedure (\ref{cg},\ref{ea}) works for any $t>-\infty$.

However, we only know $P^{\Phi_{a}}$, an incomplete specification of the
initial state of $\mathcal{S}$\cite{tol}, so that a wave function description
for $\mathcal{S}$ cannot last very long. Let us use%
\begin{equation}
P^{\Phi_{a}}|1+\int_{-\infty}^{t}dt^{\prime}\int d\tau\Phi_{a}^{\ast}%
[-\frac{i}{\hbar}H_{1}(t^{\prime})]\Phi_{a}|\label{coh}%
\end{equation}%
\[
\thicksim\sum_{b(\neq a)}P^{\Phi_{b}}|\int_{-\infty}^{t}dt^{\prime
}e^{it^{\prime}\omega_{ab}}\int d\tau\Phi_{a}^{\ast}[-\frac{i}{\hbar}%
H_{1}(t^{\prime})]\Phi_{b}|
\]
to define a coherence time $t_{a}^{coh}$ for state $\Psi_{a}$. The expression
inside the modulus sign on the LHS is the remaining probability amplitude of
state $\Phi_{a}$ at time $t$ if $\Psi_{a}(t=-\infty)=\Phi_{a}$, the expression
inside the modulus sign on the RHS is the probability amplitude of state
$\Phi_{a}$ at time $t$ if $\Psi_{b}(t=-\infty)=\Phi_{b}$. When $t>t_{\max
}^{coh}=\max\{t_{a}^{coh},\forall a\}$, it is not possible to describe
$\mathcal{S}$ with a wave function. The energy exchange between $\mathcal{S}$
and $\mathcal{B}$ makes the system a member of the canonical ensemble at
temperature $T$, and the system distributes itself into various stationary
states $\{\Phi_{a}\}$ according to Eq.(\ref{canp}). The incomplete
specification of initial state (the indeterminacy of initial state) makes the
system dephase in a time period $t_{\max}^{coh}$. The phase randomization
continues ceaselessly, so that one no longer needs the solution $\Psi_{a}(t)$
of Eq.(\ref{eu1}) for $t>t_{\max}^{coh}$.

Expressions (\ref{z1},\ref{fc}) indicate that Dirac's perturbation theory can
be used only for a time interval much shorter than
\begin{equation}
t_{p}=\frac{\hbar}{(H_{1})_{ba}}.\label{prt}%
\end{equation}
We can see that $t_{p}$ is the same order as $t_{\max}^{coh}$. In other words,
the method of variation of constants fails long before\ the system loses its
wave function description. The multi-scale method and Fano's ansatz\cite{fan}
can be used to obtain $\Psi_{a}(t)$ for a longer period: we will discuss them
elsewhere. As long as the system can be described by a wave function
($t<t_{\max}^{coh}$), we can apply the method of variation of constants to
calculate the time evolution of wave function.

\subsubsection{Error arising from the indeterminacy of the interaction time}

We show that the indeterminacy of the interaction time will not lead to any
serious error in the results obtained by the MVC.\

From%
\begin{equation}
\frac{dg_{ba}^{(1)}}{dt}=-\frac{i}{\hbar}e^{i(E_{b}-E_{a})t/\hbar}\int
d\tau\Phi_{b}^{\ast}H_{1}(t)\Phi_{a}, \label{p1}%
\end{equation}
we see that the characteristic interaction time is $(\omega_{ba}+\omega)^{-1}%
$, where $\omega$ is the frequency of external conditions. For some
interaction time $t$,
\begin{equation}
g_{ba}^{(1)}(t)\thicksim\frac{e^{it(\omega_{ba}+\omega)}}{\hbar(\omega
_{ba}+\omega)}\int d\tau\Phi_{b}^{\ast}H_{1}\Phi_{a}. \label{1cg}%
\end{equation}

Because any external disturbance (including velocity gradient) is only exerted
on a few degrees of freedom, the external distance can quickly relax to other
degrees of freedom\cite{fan}. There is an indeterminacy $\Delta t$ in the
interaction time, which is the order of microscopic relaxation time
$t_{m}\thicksim\hbar/U$. The error caused by this indeterminacy is%
\begin{equation}
\Delta g_{ba}^{(1)}(t)=\frac{dg_{ba}^{(1)}}{dt}\Delta t\thicksim
-ie^{i(E_{b}-E_{a})t/\hbar}\frac{\int d\tau\Phi_{b}^{\ast}H_{1}\Phi_{a}}{U}.
\label{er}%
\end{equation}
The relative error is
\begin{equation}
\frac{\Delta g_{ba}^{(1)}(t)}{g_{ba}^{(1)}(t)}\thicksim\frac{\hbar(\omega
_{ba}+\omega)}{U}<<1. \label{re}%
\end{equation}
In addition, the phase randomization induced by the energy exchange between
bath and system leads to a destructive interference in any observable which is
bilinear in wave function $\Psi_{a}(t)$, i.e., the error will not accumulate
with time. Therefore the indeterminacy in the interaction time does not cause
serious error in any observable.

This conclusion will not change for two degenerate levels $E_{b}=E_{a}$ in a
static field $\omega=0$ and for two levels that are in resonance with external
condition $\omega_{ba}+\omega=0$. In both situations, the characteristic
interaction time is $\hbar\lbrack\int d\tau\Phi_{b}^{\ast}H_{1}\Phi_{a}]^{-1}%
$, the relative error of transition amplitude is $U^{-1}\int d\tau\Phi
_{b}^{\ast}H_{1}\Phi_{a}<<1$. For details, see Ref.\cite{deg}.

\subsection{Self-consistency}

We show that the MVC is compatible with spatial coarse-graining (\ref{cg}).
The effect of a monochromatic external field is simple\cite{pss}. An
electromagnetic field or a gravitational field, affects every molecule at the
same time. The whole system reaches steady state after a microscopic
relaxation time $\hbar/U$. The cutoff wave vector $k_{0}$ is determined by the
precision of macroscopic measurement, is not related to any time scale. But
for a velocity gradient, the length scale of spatial coarse-graining relates
to an intrinsic time scale.

\subsubsection{Monochromatic velocity gradient and external field}

If a velocity gradient with time dependence $\cos\omega t$ is applied to
$\mathcal{S}$ by a plate or a rod, after a time period several times of the
characteristic time $t_{rf}=\max\{(\omega_{ba}+\omega)^{-1},\hbar\lbrack\int
d\tau\Phi_{b}^{\ast}H_{1}\Phi_{a}]^{-1}\}$, the molecules within a distance
$t_{rf}\hbar/(md)$ to the plate or rod reach steady state. $g_{ba}^{(1)}(t)$
and momentum flux $\Pi_{\gamma\beta}(\mathbf{r},t)$ contain terms with
$\cos\omega t$ factor and terms with $\sin\omega t$ factor. The coefficient of
$\cos\omega t$ gives us the real part of the viscosity $\operatorname{Re}%
\eta(\omega)$, the coefficient of $\sin\omega t$ gives us the imaginary part
of the viscosity $\operatorname{Im}\eta(\omega)$. For the aim of computing
viscosity, we choose cutoff wave vector $k_{0}$ through $k_{0}^{-1}%
=t_{rf}\hbar/(md)$.

For the steady state under a monochromatic velocity gradient, the macroscopic
response of $\mathcal{S}$ at time $t$ corresponds to the microscopic response
at the same time $t$. The averaging procedure (\ref{cg},\ref{ea}) reflects
this fact.

\subsubsection{Disturbance with arbitrary time dependence}

\label{caus}

For a time-dependent \textit{mechanical} perturbation, the macroscopic
response at time $t$ depends on all the past history $(-\infty,t)$ of the
external conditions. Therefore the macroscopic response is a convolution of
all monochromatic transport coefficients and the external
disturbance\cite{kub}. For a velocity gradient with arbitrary time-dependence,
the macroscopic response is more complicated.

For an internal disturbance with arbitrary time dependence, we make a temporal
Fourier resolution:
\begin{equation}
H_{1}(t)=\int_{-\infty}^{\infty}d\omega H_{1\omega}e^{i\omega t}. \label{fa}%
\end{equation}
Because $H_{1}(t)$ in Eq.(\ref{eu1}) is complex, we will not require
$H_{1}(\omega)=[H_{1}(-\omega)]^{\ast}$. The first order transition amplitude
is a sum of every monochromatic transition amplitude:%
\begin{equation}
g_{ba}^{(1)}(t)=\int_{-\infty}^{\infty}d\omega\lbrack-\frac{i}{\hbar}%
\int_{-\infty}^{t}dt^{\prime}e^{i(\omega+\omega_{ba})t^{\prime}}\int d\tau
\Phi_{b}^{\ast}H_{1\omega}\Phi_{a}]. \label{sr}%
\end{equation}
According to Eqs.(\ref{fl},\ref{mdu}), the macroscopic momentum flux is a
frequency integral over all monochromatic momentum flux. One must notice that
this result originated from \textit{first} order perturbation theory. Unlike a
classical oscillator, in Eq.(\ref{eu1}) the external conditions are multiplied
by $\Psi_{a}(t)$. Although Eq.(\ref{eu1}) is linear in $\Psi_{a}$, the
response to $H_{1}(t)$ is not the sum of the responses to $\{H_{1\omega}\}$ in general.

\section{Non-equilibrium density matrix}

\label{fph}

\subsection{Equivalence between two definitions of density matrix}

The density matrix of the system is often defined as\cite{tol}%
\begin{equation}
\rho_{nm}=\frac{1}{\mathcal{M}}\sum_{\xi=1}^{\mathcal{M}}a_{(\xi)}^{\ast
}(m,t)a_{(\xi)}(n,t),\label{wd}%
\end{equation}
where $\mathcal{M}$ is the number of systems in an ensemble, $a_{(\xi)}(n,t)$
is the probability amplitude that the $\xi^{th}$ system is in state $\Phi_{n}$
at time $t$:%
\begin{equation}
\Psi(t)=\sum_{l}a(l,t)\Phi_{l}.\label{exw}%
\end{equation}
The density matrix $\widetilde{\rho}$ can also be defined by the initial value
problem\cite{kub}:%
\begin{equation}
i\hbar\frac{\partial\widetilde{\rho}}{\partial t}=[H_{0}+H_{1}(t),\widetilde
{\rho}],\text{ }\widetilde{\rho}(t=-\infty)=\rho^{eq}.\label{liu}%
\end{equation}
Here $\rho^{eq}=e^{-\beta H_{0}}/Z$ is the equilibrium density operator, and
$Z=$Tr$e^{-\beta H_{0}}$. By means of%
\begin{equation}
i\hbar\frac{\partial a_{p}(t)}{\partial t}=a_{p}(t)E_{p}+\sum_{l}%
a_{l}(t)\langle p|H_{1}(t)|l\rangle,\label{ase}%
\end{equation}
it is easy to check that if $H_{1}(t)$ is hermitian, $\widetilde{\rho}$ is the
same as $\rho$ in Eq.(\ref{wd}).

Integrating by parts, $H_{d}$ becomes
\[
H_{d}=-\frac{1}{2}P^{\Phi_{a}}\sum_{\gamma\sigma\mu_{\sigma}\nu_{\sigma}%
}\lambda_{\mu_{\sigma}\nu_{\sigma}}^{\gamma(\sigma)}\int d\mathbf{R}%
f(\mathbf{r}_{\mu_{\sigma}\nu_{\sigma}}^{(\sigma)}-\mathbf{R})
\]%
\begin{equation}
\times\sum_{j=1}^{N}\{\delta(\mathbf{r}_{j}-\mathbf{R})p_{j\gamma}+p_{j\gamma
}\delta(\mathbf{r}_{j}-\mathbf{R})\}, \label{dh}%
\end{equation}
where $p_{j\gamma}=-i\hbar\partial/\partial r_{j\gamma}$ ($\gamma=x,y,z$) is
the momentum operator for the $j^{th}$ particle. We see that $H_{1}%
(t)=H_{f}(t)+H_{c}(t)+H_{d}$ is hermitian. Thus we can use either
Eq.(\ref{wd}) or Eq.(\ref{liu}) to define density matrix. The former is more
convenient for us since we already obtained probability amplitudes in
Sec.\ref{cz}.

\subsection{Equivalence between procedure (\ref{cg},\ref{ea}) and average over
density matrix}

\label{dj2}

The perturbation expansion (\ref{sup1}) for a time-dependent wave function
$\Psi_{a}(t)$ leads to a similar expansion for the density matrix%
\begin{equation}
\rho_{cb}(t)=\rho_{cb}^{(0)}(t)+\rho_{cb}^{(1)}(t)+\cdots, \label{dem}%
\end{equation}
where the superscripts indicate the order in $H_{1}(t)$. Because (i) the
system is enclosed by a bath with temperature $T$; and (ii) a state with
initial value $\Phi_{a}$ is described by Eq.(\ref{sup1}). The zero order
density matrix is\cite{tol,cb} $\rho_{cb}^{(0)}(t)=P^{\Phi_{b}}\delta_{bc}$.
It is easy to check that $\rho_{bb}^{(1)}(t)=0$, and the first order
correction for off-diagonal element is%
\begin{equation}
\rho_{cb}^{(1)}(t)=P^{\Phi_{b}}c_{cb}^{(1)}(t)+P^{\Phi_{c}}c_{bc}^{\ast
(1)}(t). \label{1dm}%
\end{equation}
In the basis \{$\Phi_{a}$\}, the matrix element of the momentum flux operator
(\ref{dfu}) is%
\begin{equation}
\lbrack\widehat{\Pi}_{\alpha\beta}(\mathbf{r})]_{bc}=\langle\Phi_{b}%
|\widehat{\Pi}_{\alpha\beta}(\mathbf{r})|\Phi_{c}\rangle=\frac{\hbar^{2}}%
{2m}N\int d\tau_{1} \label{ymf}%
\end{equation}%
\[
\lbrack\frac{\partial\Phi_{c}}{\partial r_{\alpha}}\frac{\partial\Phi
_{b}^{\ast}}{\partial r_{\beta}}+\frac{\partial\Phi_{b}^{\ast}}{\partial
r_{\alpha}}\frac{\partial\Phi_{c}}{\partial r_{\beta}}-\frac{1}{2}%
\frac{\partial^{2}(\Phi_{c}\Phi_{b}^{\ast})}{\partial r_{\beta}\partial
r_{\alpha}}]
\]%
\[
-\frac{1}{2}N\int d\tau_{1}\sum_{k=2}^{N}(\mathbf{r}_{\beta}-\mathbf{r}%
_{k\beta})\frac{\partial U(\mathbf{r}_{k},\mathbf{r})}{\partial r_{\alpha}%
}\Phi_{c}\Phi_{b}^{\ast}.
\]
According to standard statistical mechanics\cite{kub,tol,cb}, the observed
macroscopic momentum flux $\Pi_{\alpha\beta}(\mathbf{r},t)$ is%
\begin{equation}
\Pi_{\alpha\beta}(\mathbf{r},t)=\int d\mathbf{R}f(\mathbf{r}-\mathbf{R}%
)\sum_{bc}[\widehat{\Pi}_{\alpha\beta}(\mathbf{R})]_{bc}\rho_{cb}^{(1)}(t).
\label{usa}%
\end{equation}
Substitute Eq.(\ref{ymf}) into Eq.(\ref{usa}), we obtain Eq.(\ref{mrf}) which
was obtained by the microscopic response method.

\subsection{Constancy of the entropy $S^{\mathcal{S}}$ of system $\mathcal{S}%
$}

The entropy $S^{\mathcal{S}}$ of system $\mathcal{S}$ is defined
by\cite{tol,kub,cb}%
\begin{equation}
S^{\mathcal{S}}(t)=-\text{Tr}\rho\ln\rho=-\sum_{n}\langle n|\rho\ln
\rho|n\rangle, \label{shs}%
\end{equation}
where $\langle\mathbf{r}_{1}\cdots\mathbf{r}_{N}|n\rangle=\Phi_{n}$ is the
eigenfunction of $H_{0}$. The time dependence of entropy is included in the
density operator $\rho(t)$. The time rate of change of $S^{\mathcal{S}}(t)$ is
written as%
\begin{equation}
\frac{\partial S^{\mathcal{S}}(t)}{\partial t}=-\sum_{n}\langle n|\frac
{\partial\rho\ln\rho}{\partial t}|n\rangle. \label{sb}%
\end{equation}

Let $A=1-\rho$, then $\rho=1-A$. It is easy to check that%
\begin{equation}
i\hbar\frac{\partial A^{n}}{\partial t}=[H_{0}+H_{1},A^{n}],\text{
\ }n=1,2,3,\cdots. \label{na}%
\end{equation}
Making use of%
\[
\ln\rho=\ln(1-A)=-A-\frac{A^{2}}{2}-\frac{A^{3}}{3}-\cdots,
\]
Eq.(\ref{na}) leads to a well-known result\cite{zub}%

\begin{equation}
i\hbar\frac{\partial\ln\rho}{\partial t}=[H_{0}+H_{1}(t),\ln\rho]. \label{ze}%
\end{equation}
Then, we combine Liouville equation (\ref{liu}) and Eq.(\ref{ze}), to obtain%
\begin{equation}
i\hbar\frac{\partial\rho\ln\rho}{\partial t}=[H_{0}+H_{1}(t),\rho\ln\rho].
\label{seq}%
\end{equation}

By means of Eq.(\ref{seq}), Eq.(\ref{sb}) becomes%

\begin{equation}
\frac{\partial S^{\mathcal{S}}(t)}{\partial t}=\frac{i}{\hbar}\sum_{n}\langle
n|[H_{0}+H_{1}(t),\rho\ln\rho]|n\rangle=0. \label{zss}%
\end{equation}
If a system is in good thermal contact with a reservoir such that the heat
generated by the external field and velocity gradient can be instantaneously
transferred to the surrounding bath, its temporal evolution can be described
Eq.(\ref{eu1}) or Eq.(\ref{liu}). The entropy $S^{\mathcal{S}}$ of system
$\mathcal{S}$ is a constant. The irreversibility (entropy production) of the
momentum transport process is reflected in Eqs.(\ref{entr},\ref{tre}).

Eq.(\ref{zss}) does not depend on the initial condition for $\rho$. The key
reasons are that (1) the system is described by a `Hamiltonian'. The density
matrix satisfies a memoryless Liouville equation (\ref{liu}); (2)
Tr$AB=$Tr$BA$ is correct for two arbitrary operators $A$ and $B$. For the same
reasons, if the entropy of a microscopic canonical ensemble is defined by
Eq.(\ref{shs}), one has also\cite{cb} $\partial S^{\mathcal{S}}(t)/\partial
t=0$. Removing irrelevant degrees of freedom can avoid this
difficulty\cite{zw61,sew,mo}.

\subsection{Cumulant expansion: connection with previous methods\ }

Eqs.(\ref{liu},\ref{ze},\ref{seq}) have a common form%
\begin{equation}
i\hbar\frac{\partial D}{\partial t}=[H_{0}+H_{1}(t),D], \label{ce}%
\end{equation}
where $D=\rho$, $\ln\rho$ and $\rho\ln\rho$. One can solve Eq.(\ref{ce}) with
perturbation method\cite{kub,cb}%
\begin{equation}
D(t)=D^{(0)}(t)+D^{(1)}(t)+D^{(2)}(t)+\cdots, \label{pte}%
\end{equation}
where the superscript indicates the order in $H_{1}(t)$. Before we introduce
external field and velocity gradient, the system is in equilibrium. The
initial value of $D(t)$ is its equilibrium value $D^{eq}$: $D^{(0)}%
(-\infty)=D^{eq}$ and $D^{(q)}(-\infty)=0$, $q=1,2,\cdots$. Since $D^{eq}$ is
function of $H_{0}$, the zero order solution of Eq.(\ref{ce}) is\cite{kub}%
\begin{equation}
D^{(0)}(t)=e^{-i(t-t^{\prime})H_{0}/\hbar}D^{(0)}(t^{\prime})e^{i(t-t^{\prime
})H_{0}/\hbar}=D^{eq}. \label{0s}%
\end{equation}
The first order solution is\cite{kub}%
\begin{equation}
D^{(1)}(t)=-\frac{i}{\hbar}\int_{-\infty}^{t}dt^{\prime}e^{-i(t-t^{\prime
})H_{0}/\hbar}[H_{1}(t^{\prime}),D^{eq}]e^{i(t-t^{\prime})H_{0}/\hbar}.
\label{1s}%
\end{equation}
It is easy to check that:%
\begin{equation}
\langle n|D^{(1)}(t)|n\rangle=0 \label{d0}%
\end{equation}
for any eigenstate $\Phi_{n}$ of $H_{0}$.

For $D=\ln\rho$, $[\ln\rho](t=-\infty)=\ln\rho^{eq}$, so that
\begin{equation}
\lbrack\ln\rho]^{(0)}(t)=-\beta H_{0}-\ln Z. \label{xz0}%
\end{equation}
The first order solution is%
\begin{equation}
\lbrack\ln\rho]^{(1)}(t)=\frac{i\beta}{\hbar}\int_{-\infty}^{t}dt^{\prime
}e^{-i(t-t^{\prime})H_{0}/\hbar}[H_{1}(t^{\prime}),H_{0}]e^{i(t-t^{\prime
})H_{0}/\hbar}. \label{1he}%
\end{equation}
We recognize that $[H_{1}(t^{\prime}),H_{0}]/\hbar$ is the dissipated power
caused by the external field and velocity gradient. Excepting a local
non-uniformity of temperature which does not appear in the canonical ensemble,
$[\ln\rho]^{(1)}(t)$ is Mori's $\overset{\cdot}{R}$\cite{ori}.To obtain the
term linear in velocity gradient $v_{x0}/l_{y}$, we only need $[H_{d}%
,H_{0}]/\hbar$. By means of Eqs.(\ref{hs},\ref{eu4}), the term linear in
$v_{x0}/l_{y}$ is%
\[
\langle b|[\ln\rho]^{(1)}(t)|a\rangle=-P^{\Phi_{a}}\beta\int_{-\infty}%
^{t}dt^{\prime}\sum_{\gamma\sigma\mu_{\sigma}\nu_{\sigma}}\lambda_{\mu
_{\sigma}\nu_{\sigma}}^{\gamma(\sigma)}%
\]%
\[
\int d\mathbf{R}f(\mathbf{r}_{\mu_{\sigma}\nu_{\sigma}}^{(\sigma)}%
-\mathbf{R})\langle b|e^{-i(t-t^{\prime})H_{0}/\hbar}%
\]%
\[
\{\sum_{jl(l\neq j)}\delta(\mathbf{r}_{j}-\mathbf{R})\frac{\partial
U(\mathbf{r}_{j},\mathbf{r}_{l})}{\partial r_{j\gamma}}+\frac{\hbar^{2}}%
{m}\sum_{j}[\nabla_{\mathbf{r}_{j}}\delta(\mathbf{r}_{j}-\mathbf{R}%
)]\cdot\nabla_{\mathbf{r}_{j}}\frac{\partial}{\partial r_{j\gamma}}%
\]%
\[
+\frac{\hbar^{2}}{2m}\sum_{j}[\nabla_{\mathbf{r}_{j}}^{2}\delta(\mathbf{r}%
_{j}-\mathbf{R})]\frac{\partial}{\partial r_{j\gamma}}+\frac{\hbar^{2}}%
{2m}\sum_{j}[\nabla_{\mathbf{r}_{j}}\frac{\partial\delta(\mathbf{r}%
_{j}-\mathbf{R})}{\partial r_{j\gamma}}]\cdot\nabla_{\mathbf{r}_{j}}%
\]%
\begin{equation}
+\frac{\hbar^{2}}{4m}\sum_{j}[\nabla_{\mathbf{r}_{j}}^{2}\frac{\partial
\delta(\mathbf{r}_{j}-\mathbf{R})}{\partial r_{j\gamma}}]\}e^{i(t-t^{\prime
})H_{0}/\hbar}|a\rangle. \label{xz1}%
\end{equation}
Eq.(\ref{xz1}) is $\beta$ multiplied by the dissipated energy from $t=-\infty$
to $t$.

Eqs.(\ref{xz0},\ref{xz1}) forms a cumulant expansion for the density matrix
$\rho=e^{[\ln\rho]}$:%
\begin{equation}
\rho=Z^{-1}\exp\{-\beta H_{0}+[\ln\rho]^{(1)}(t)+\cdots\}.\label{czk}%
\end{equation}
The operator in \{\} on RHS of Eq.(\ref{xz1}) is the divergence of the stress
tensor (momentum flux). If we replace the momentum flux operator with its
expectation value, and change the source of velocity gradient from boundary
surfaces to bulk, Eq.(\ref{czk}) is reduced to the non-equilibrium density
matrix obtained by many different methods\cite{kyn,mac1,mac2,zub}. We need not
invoke Onsager's regression assumption\cite{kyn} and nonconservative force
from bath\cite{mac1,mac2}.

\section{Summary}

\label{smy}

According to the microscopic response method\cite{short,pss,jp}, the observed
macroscopic momentum density Eq.(\ref{ond}) in a many-body system can be
obtained from the microscopic momentum density in a pure state by spatial
coarse-graining and averaging over all possible initial conditions. If we
adopt a no-slip boundary condition Eq.(\ref{nsli}), we can view a velocity
gradient as a constraint on the many-body wave function of the system
Eq.(\ref{bc}). The evolution equation for wave function Eq.(\ref{eu1}) is then
derived from the Lagrange multiplier method. The Lagrange multipliers have
been obtained by the method of variation of constants, Eqs.(\ref{Li}%
,\ref{L1},\ref{L2}). They express the interactions on the system exerted by
the moving boundary plate and three static plates.

From the evolution equation Eq.(\ref{eu1}), we calculated the time rate of
change of mass density, momentum density and total energy in a pure state:
Eqs.(\ref{hmc0},\ref{cond},\ref{tre}). The dissipation caused by a velocity
gradient contains $[P^{\Phi_{a}}]^{2}$, in contrast to the dissipation caused
by a mechanical disturbance\cite{jp} which contains $P^{\Phi_{a}}$. This is an
obviously statistical character, velocity gradient is an internal disturbance.

Although velocity gradient is an internal disturbance, by means of the
conservation law of momentum in a pure state, we could still define momentum
flux in a pure state Eq.(\ref{flu}). The observed macroscopic momentum flux
Eq.(\ref{mdu}) is obtained by spatial coarse-graining the momentum flux in a
pure state and averaging over all possible initial conditions. The shear
viscosity can be read out from Eq.(\ref{mrf}). Comparing to the traditional
theories of viscosity, the present \textit{ansatz} does not need temporal
coarse graining. Taking a spatial coarse-grained average over the microscopic
response automatically contains temporal coarse-graining and coarse-graining
in the eigenvalue spectrum of collective variables. We compared the
Hamiltonian derived from Lagrange multiplier method with the phenomenological
non-equilibrium Hamiltonian. The non-equilibrium density matrix implied by the
present method can be reduced to those derived by other methods.

The present \textit{ansatz} can be generalized to all internal disturbances.
We can view concentration gradient, temperature gradient and velocity gradient
as constraints on the many-body wave function of system, and have an unified
theory for diffusion, thermal conductivity and viscosity. This work is in progress.

\begin{acknowledgements}
This work is supported by the Army Research Laboratory and Army Research
Office under Grant No. W911NF1110358 and the NSF under Grant DMR 09-03225.
\end{acknowledgements}

\end{document}